\documentclass[useAMS,usenatbib]{mn2e}
\usepackage{listings}
\usepackage{graphics}
\usepackage{subfigure}
\usepackage{graphicx}
\usepackage{amssymb}
\usepackage{hyperref}
\usepackage{deluxetable}
\usepackage{setspace}
\usepackage{multirow}
\usepackage{slashbox}

\newcommand{\lya}{Ly-$\alpha$}
\newcommand{\hi}{H~{\sc i}}
\newcommand{\ovi}{O~{\sc vi}}
\newcommand{\civ}{C~{\sc iv}}
\newcommand{\nv}{N~{\sc v}}
\newcommand{\ciii}{C~{\sc iii}}
\newcommand{\siiv}{Si~{\sc iv}}
\newcommand{\siiii}{Si~{\sc iii}}
\newcommand{\feiii}{Fe~{\sc iii}}

\newcommand{\kms}{km s$^{-1}$}
\newcommand{\rvoid}{$R_{\rm{void}}$}
\newcommand{\nhi}{$N_{\rm HI}$}
\newcommand{\bhi}{$b_{\rm HI}$}
\newcommand{\mpc}{$h^{-1}$Mpc}
\newcommand{\cm}{cm$^{-2}$}
\newcommand{\fn}{}
\newcommand{\fnn}{\footnotemark[11]}
\newcommand{\gimic}{{\sc gimic}}

\title[LSS in absorption: Gas within and around Voids]{Large Scale
  Structure in Absorption: Gas within and around Galaxy Voids}
\author[Nicolas Tejos et al.]{
\parbox[t]{\textwidth}{
\vspace{-1.0cm}
Nicolas Tejos$^{1}$\thanks{E-mail: nicolas.tejos@durham.ac.uk},
Simon L. Morris$^{1}$, 
Neil H. M. Crighton$^{2}$, 
Tom Theuns$^{1}$,
Gabriel Altay$^{1}$, 
Charles W. Finn$^{1}$
}
\vspace*{6pt} \\
$^{1}$Department of Physics, Durham University, South Road, Durham, DH1 3LE, UK.\\
$^{2}$Max Planck Institute for Astronomy, K{\"o}nigstuhl 17, D-69117 Heidelberg, Germany.
\vspace*{-0.5cm}}

\begin{document}
\date{Draft version}

\pagerange{\pageref{firstpage}--\pageref{lastpage}} \pubyear{2012}

\maketitle

\label{firstpage}

\begin{abstract}
We investigate the properties of the \hi~\lya~absorption systems
(\lya~forest) {\it within} and {\it around} galaxy voids at
$z\lesssim0.1$. We find a significant excess ($>99$ per cent confidence
level, c.l.)  of \lya~systems at the edges of galaxy voids with respect
to a random distribution, on $\sim 5$ \mpc~scales. We find no
significant difference in the number of systems inside voids with
respect to the random expectation. We report differences between both
column density (\nhi) and Doppler parameter (\bhi) distributions of
\lya~systems found inside and at the edge of galaxy voids at the
$\gtrsim98$ and $\gtrsim90$ per cent c.l. respectively. Low density
environments (voids) have smaller values for both \nhi~and \bhi~than
higher density ones (edges of voids). These trends are theoretically
expected and also found in \gimic, a state-of-the-art hydrodynamical
simulation. Our findings are consistent with a scenario of {\it at
  least} three types of \lya~systems: (1) containing embedded galaxies
and so directly correlated with galaxies (referred as `halo-like'), (2)
correlated with galaxies only because they lie in the same over-dense
LSS, and (3) associated with under-dense LSS with a very low
auto-correlation amplitude ($\approx$ random) that are not correlated
with luminous galaxies. We argue the latter arise in structures still
growing linearly from the primordial density fluctuations inside galaxy
voids that have not formed galaxies because of their low densities. We
estimate that these under-dense LSS absorbers account for $25-30 \pm 6$
per cent of the current \lya~population (\nhi~$\gtrsim 10^{12.5}$ \cm)
while the other two types account for the remaining $70-75 \pm 12$ per
cent. Assuming that {\it only} \nhi~$\ge 10^{14}$ \cm~systems have
embedded galaxies nearby, we have estimated the contribution of the
`halo-like' \lya~population to be $\approx 12-15 \pm 4$ per cent and
consequently $\approx 55-60 \pm 13$ per cent of the \lya~systems to be
associated with the over-dense LSS.
\end{abstract}

\begin{keywords}
intergalactic medium: \lya~forest --large-scale structure of the
universe: galaxy voids, galaxy filaments --quasars: absorption lines
\end{keywords}

\section{Introduction}

The inter-galactic medium (IGM) hosts the main reservoirs of baryons at
all epochs (see \citealt{prochaska09} for a review). This is supported
by both observations \citep[e.g.,][]{fukugita98,fukugita04,shull11} and
simulations \citep[e.g.,][]{cen99,theuns99,dave10}. Efficient feedback
mechanisms that expel material from galaxies to the IGM are required to
explain the statistical properties of the observed galaxies
\citep[e.g.,][]{baugh05,bower06,schaye10}. Given that galaxies are
formed by accreting gas from the IGM, a continuous interplay between
the IGM and galaxies is then in place. Consequently, understanding the
relationship between the IGM and galaxies is key to understanding
galaxy formation and evolution. This has been recognized since the
earliest {\it Hubble Space Telescope } (HST) spectroscopy of QSOs,
where the association between low-$z$ IGM absorption systems and
galaxies was investigated for the first time
\citep[e.g.,][]{spinrad93,morris93,morris94,stocke95,lanzetta95}.

The large scale environment in which matter resides is also important,
as it is predicted \citep[e.g.,][]{borgani02,padilla09} and observed
\citep[e.g.,][]{lewis02,lopez08,padilla10} to have non negligible
effects on the gas and galaxy properties. Given that baryonic matter is
expected to fall into the considerably deeper gravitational potentials
of dark matter, the IGM gas and galaxies are expected to be
predominantly found at such locations forming the so called `cosmic
web' \citep{bond96}. Identification of large scale structures (LSS)
like galaxy clusters, filaments or voids and their influence over the
IGM and galaxies is then fundamental to a complete picture of the
IGM/galaxy connection and its evolution over cosmic time.

With the advent of big galaxy surveys such as the {\it 2dF}
\citep{colless01} or the {\it Sloan Digital Sky Survey}
\citep[SDSS,][]{abazajian09} it has been possible to directly observe
the nature and extent of the distribution of stellar matter in the
local universe. Galaxies tend to lie in the filamentary structure which
simulations predict, however, very little is known about the {\it
  actual} gas distribution at low-$z$. In this work we focus on the
study of \hi~\lya~(hereafter referred simply as \lya) absorption
systems found {\it within} and {\it around} galaxy voids at $z \lesssim
0.1$.

Galaxy voids are the best candidates to start our statistical study of
LSS in absorption. Voids account for up to $60-80\%$ of the volume of
the universe at $z=0$ \citep[e.g.,][]{aragon-calvo10,pan11}. Some
studies have suggested that when a minimum density threshold is
reached, voids grow in a spherically symmetric way
\citep[e.g.,][]{regos91,vandeweygaert93}. This suggests that voids have
a relatively simple geometry, which makes them comparatively easy to
define and identify from current galaxy surveys \citep[although
  see][for a discussion on different void finder
  algorithms]{colberg08}. Galaxy voids are a unique environment in
which to look for evidence of early (or even primordial) enrichment of
the IGM \citep[e.g.,][]{stocke07}. It is interesting that galaxy voids
are present even in the distribution of low mass galaxies
\citep[e.g.,][]{peebles01,tikhonov09} and so there must be mechanisms
that prevent galaxies from forming in such low density environments.

Previous studies of \lya~absorption systems associated with voids at
low-$z$ have relied on a `nearest galaxy distance' (NGD) definition.
\citep[e.g.,][]{penton02,stocke07,wakker09}. In order to have a clean
definition of void absorbers the NGD must be large, leading to small
samples. For instance, \citet{penton02} found only 8 void absorbers
(from a total of $46$ systems) defined as being located at $> 3$
$h^{-1}_{70}$ Mpc from the nearest $\ge L^*$ galaxy. \citet{wakker09}
found 17 void absorbers (from a total of $102$) based on the same
definition. \citet{stocke07} had to relax the previous limit to $> 1.4$
$h^{-1}_{70}$ Mpc in order to find $61$ void absorbers (from a total of
$651$ systems), although only 12 were used in their study on void
metallicities. Note that a low NGD limit (of $1.4$ $h^{-1}_{70}$ Mpc)
could introduce some contamination of not-void absorbers. This is
because filaments in the `cosmic web' are expected to be a couple of
Mpc in radius \citep{gonzalez10, aragon-calvo10, bond10}. Considering
the Local Group as an example, being $1.4$ $h^{-1}_{70}$ Mpc away from
either the Milky Way or Andromeda cannot be considered as being in a
galaxy void. On the other hand, given that there is a population of
galaxies inside voids \citep[e.g.,][]{rojas05,park07,kreckel11}, the
NGD definition could also miss some `real' void absorbers relatively
close to bright isolated galaxies. In fact, \citet{wakker09} found that
there may be no void absorbers in their sample (based on the NGD
definition) if the luminosity limit to the closest galaxy is reduced to
$0.1L^*$. Note however that their sample is very local ($z\le 0.017$ or
$\lesssim 70$ $h^{-1}_{70}$ Mpc away), and it might be biased because
of the local overdensity to which our Local Group belongs.

In this work we use a different approach to define void absorption
systems. We based our definition on current galaxy void catalogs
(typical radius of $> 14$ $h^{-1}_{70}$ Mpc), defining void absorbers
as those located inside such galaxy voids. This leads to larger samples
of well identified void absorbers compared to previous
studies. Moreover, this approach allows us to define a sample of
absorbers located at the very edges of voids, that can be associated
with walls, filaments and nodes, allowing us to get some insights in
the distribution of gas in the `cosmic web' itself. This definition is
different from the NGD based ones and it focuses on the `large scale'
($\gtrsim10$ Mpc) relationship between \lya~forest systems and
galaxies. The results from this work will offer a good complement to
previous studies based on `local' scales ($\lesssim 2$ Mpc).

Our paper is structured as follows. The catalogs of both \lya~systems
and galaxy voids that we used in this work are described in \S
\ref{samples}. Definition of our LSS in absorption samples and the
observational results are presented in \S \ref{analysis}. We compare
our observational results with a recent cosmological hydrodynamical
simulation in \S \ref{gimic}. We discuss our findings in \S
\ref{discussion} and summarize them in \S \ref{summary}. A check for
systematic effects and biases that could be present in our data
analysis is presented in Appendix \ref{systematic}. All distances are
in co-moving coordinates assuming $H_0=100 \ h$ \kms Mpc$^{-1}$,
$h=0.71$, $\Omega_{\rm m}=0.27$, $\Omega_{\rm \Lambda}=0.73$, $k=0$
cosmology unless otherwise stated. This cosmology was chosen to match
the one adopted by \citet{pan11} (D. Pann, private communication; see
\S \ref{samples:voids}).

\section{Data}\label{samples}

\subsection{Gas in absorption}\label{samples:gas}
We use QSO absorption line data from the \citet[][hereafter
  DS08]{danforth08} catalog, which is the largest high-resolution ($R
\equiv \frac{\Delta \lambda}{\lambda}\approx 30\,000-100\,000$),
low-$z$ IGM sample to date\footnote{We note that after this paper was
  submitted, a new pre-print by \citet{tilton12} appeared with an
  updated version of the DS08 catalog.}. Briefly, the catalog lists 651
\lya~absorption systems at $z_{\rm abs} \le 0.4$, with associated metal
lines (\ovi, \nv, \civ, \ciii, \siiv, \siiii~and \feiii; when the
spectral coverage and signal-to-noise allowed their observation), taken
from 28 AGN observed with both the {\it Space Telescope Imaging
  Spectrograph} \citep[STIS,][]{woodgate98} on the HST, and the {\it
  Far Ultraviolet Spectroscopic Explorer} \citep[FUSE,][]{moos00}. The
systems are characterized by their rest-frame equivalent widths
($W_r$), or upper limits on $W_r$, for each individual
transition. Column densities (\nhi) and Doppler parameters (\bhi) were
inferred using the apparent optical depth method
\citep[AODM,][]{savage91} and/or Voigt profile line fitting. In
particular for the \lya~transition, a curve-of-growth (COG) solution
was used when other Lyman series lines were available (see also
Appendix \ref{systematic:method}). We refer the reader to DS08 (and
references therein) for further description and discussion.

In order to identify absorbing gas associated with LSS (drawn from the
SDSS DR7), we use a subsample of the DS08 AGN sightlines that intersect
the SDSS volume (PG\,0953$+$414, Ton\,28, PG\,1116$+$215,
PG\,1211$+$143, PG\,1216$+$069, 3C\,273, Q\,1230$+$0115,
PG\,1259$+$593, NGC\,5548, Mrk\,1383 and PG\,1444$+$407; see Table
\ref{tab:1}). Despite the fact that PG\,1216$+$069 spectrum has a poor
quality, it is still possible to find strong systems in it, and so we
decided to do not exclude it from the sample (this inclusion does not
affect our results; see \S \ref{gasaroundvoids})\footnote{We note that
  \citet{chen09} have presented a \lya~absorption system list along
  PG\,1216$+$069 sightline at a better sensitivity than that of
  DS08. In order to have an homogeneous sample, we did not include this
  new data in our analysis however.}. We use the rest of the sightlines
in the DS08 catalog to derive the general properties of the average
absorber for comparison (see Appendix \ref{reference}).

\begin{table}
  \centering
  \begin{minipage}{80mm}
    \caption{IGM sightlines from DS08 that intersect the
      SDSS survey.}\label{tab:1}
    \begin{tabular}{@{}lcclr@{}}
      \hline
      Sight Line   & RA (J2000)  & Dec (J2000) & $z_{\rm AGN}$ & $S/N$\tablenotemark{a}\\
      
      \hline
 
      PG\,0953$+$414   & 09 56 52.4  & $+$41 15 22 &  0.23410  & $14$\\
      Ton\,28          & 10 04 02.5  & $+$28 55 35 &  0.32970  & $ 9$\\
      PG\,1116$+$215   & 11 19 08.6  & $+$21 19 18 &  0.17650  & $18$\\
      PG\,1211$+$143   & 12 14 17.7  & $+$14 03 13 &  0.08090  & $30$\\
      PG\,1216$+$069   & 12 19 20.9  & $+$06 38 38 &  0.33130  & $ 3$\\
      3C\,273          & 12 29 06.7  & $+$02 03 09 &  0.15834  & $35$\\
      Q\,1230$+$0115   & 12 30 50.0  & $+$01 15 23 &  0.11700  & $12$\\
      PG\,1259$+$593   & 13 01 12.9  & $+$59 02 07 &  0.47780  & $12$\\
      NGC\,5548        & 14 17 59.5  & $+$25 08 12 &  0.01718  & $13$\\
      Mrk\,1383        & 14 29 06.6  & $+$01 17 06 &  0.08647  & $16$\\
      PG\,1444$+$407   & 14 46 45.9  & $+$40 35 06 &  0.26730  & $10$\\
      \hline
      
    \end{tabular}
    \tablenotetext{a}{Median HST/STIS signal-to-noise ratio per
      two-pixel resolution element in the $1215-1340$ \AA~range
      (C. Danforth, private communication). The expected minimum
      equivalent width, $W_{\rm min}$, at a c.l. of $cl$ corresponding
      to a given $S/N$ can be estimated from $W_{\rm min}=\frac{cl
        \lambda}{R(S/N)}$, where $R$ is the spectral resolution (e.g.,
      see DS08).}
  \end{minipage}
\end{table}


In our analysis we focus on statistical comparisons of the
\hi~properties in different LSS environments. Metal systems have
smaller redshift coverage and lower number densities than
\lya~absorbers. Consequently, we do not aim to draw statistical
conclusions from them. We intend to pursue metallicity studies in
future work.

\subsection{Galaxy voids}\label{samples:voids}
We use a recently released galaxy-void catalog from SDSS DR7 galaxies
\citep[][hereafter P12]{pan11}, which is the largest galaxy-void sample
to date. Hereafter we will use the term void to mean galaxy-void unless
otherwise stated. P12 identified $\gtrsim 1000$ cosmic voids using the
\verb"VoidFinder" algorithm described by \citet{hoyle02}, with
redshifts between $0.01 \lesssim z \lesssim 0.102$. To summarize, it
first uses a nearest neighbor algorithm on a volume limited galaxy
survey. Galaxies whose third nearest neighbor distance is greater than
$6.3$ \mpc~are classified as potential void galaxies, whereas the rest
are classified as {\it wall} galaxies. Void regions are identified by
looking for maximal empty spheres embedded in the {\it wall} galaxy
sample. These individual void spheres have radii between $10<R_{\rm
  void} \lesssim 25$ \mpc, with mean radius $\langle R_{\rm void}
\rangle \approx 13$ \mpc. The minimum radius of $10$ \mpc~for the void
spheres was imposed. Only galaxies with spectroscopic redshifts were
used and therefore we expect the uncertainties in the void centers and
radii to be small ($\lesssim 1$ Mpc; we will discuss the effects of
peculiar velocities in \S \ref{gasaroundvoids}). Independent void
regions are defined by combining all the adjacent spheres that share
more than $10\%$ of their volume with another. Void galaxies are
defined as those galaxies that lie within a void region. We refer the
reader to P12 for further description and discussion.

\begin{figure*}
  \includegraphics[width=170mm]{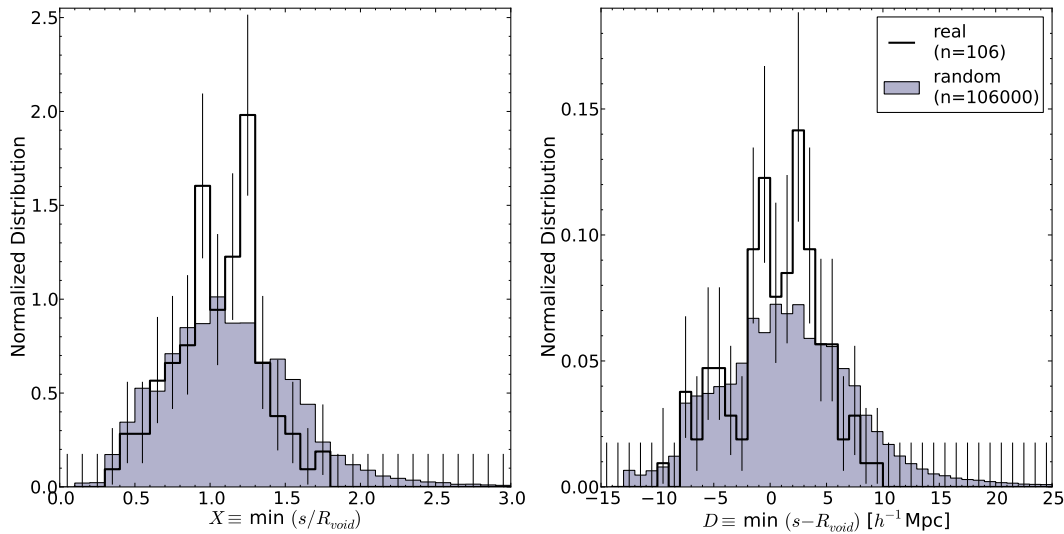}
  \caption{Normalized (in area) distribution of \hi~absorption systems
    as a function of $X$ (left panel; $0.1$ binning) and $D$ (right
    panel; $1$ $h^{-1}$Mpc binning) for both real and random
    samples. Error bars correspond to the Poissonian uncertainty from
    the analytical approximation $\sigma^+_n\approx \sqrt{n +
      \frac{3}{4}}+1$ and $\sigma^-_n\approx \sqrt{n - \frac{1}{4}}$
    \citep{gehrels86}. Real and random distributions are different at a
    $\gtrsim 99.5\%$ confidence level (see \S \ref{gasaroundvoids} for
    further details). }

  \label{fig:X_D_dist}
\end{figure*}

In our analysis, for simplicity, we use the individual spheres as
separate voids instead of using the different independent void regions.
This choice has the following advantages. First, it allows us to use a
perfectly spherical geometry, making it possible to characterize each
void by just one number: its radius. Thus, we can straightforwardly
scale voids with different sizes for comparison. Secondly, this
approach allows us to identify regions at the very edges of the
voids. P12 found that the number density of galaxies has a sharp peak
at a distance $\approx R_{\rm{void}}$ from the center of the void
spheres, a clear signature that walls are well defined (at least from
the point of view of bright galaxies at low redshifts). This is also
consistent with the predictions of linear gravitation theory
\citep[e.g.,][]{icke84,sheth04} and dark matter simulations
\citep[e.g.,][P12]{benson03,colberg05,ceccarelli06}. Therefore, by
looking for absorption systems very close to the edge of voids, we
expect to trace a different cosmic environment. Thirdly, using the
individual void spheres securely identifies void regions. The void-edge
sample on the other hand, could be contaminated by void regions
associated with the intersections of two void spheres. We checked that
this is not the case though (see \S \ref{definition}). This
contamination should only reduce any possible difference between the
two samples rather than enhance them. We also note that systematic
uncertainties produced by assuming voids to be perfect spheres should
also act to reduce any detected difference.

\section{Data analysis and results}\label{analysis}

\subsection{Number density of absorption systems around voids}\label{gasaroundvoids}
We have cross-matched the IGM absorption line catalog from DS08 (see \S
\ref{samples:gas}) with the void catalog from P12 (see \S
\ref{samples:voids}). A total of $106$ \lya~absorption systems were
found in the $11$ sightlines that intersect the void sample volume
(i.e., those with $0.01 \le z_{\rm abs} \le 0.102$).

We first look for a possible difference in the number density of
\lya~systems as a function of the distance to voids. We take two
approaches. First, we define $X$ as the three-dimensional distance
between an absorption system and the closest void center in
\rvoid~units, so

\begin{equation}
X \equiv \min_{\rm{sample}}\frac{s}{R_{\rm{void}}} \ \rm{,}
\label{eq:X}
\end{equation}

\noindent where $s$ is the co-moving distance between the center of the
closest void and the absorber. Thus, $0 \le X <1$ corresponds to
absorption systems inside voids and $X > 1$ corresponds to absorption
systems outside voids. A value of $X \approx 1$ corresponds to
absorption systems around void edges as defined by the galaxy
distribution.

The second approach defines $D$ as the three-dimensional distance
between an absorption system and the closest void edge in co-moving
\mpc, so

\begin{equation}
D \equiv \min_{\rm{sample}} (s-R_{\rm{void}}) \ [h^{-1}\rm{Mpc].}
\label{eq:D}
\end{equation}

\noindent Negative $D$ values correspond to absorption systems inside
voids while positive values correspond to absorbers outside
voids. Values of $D\approx 0$ \mpc~are associated with absorption
systems around void edges as defined by the galaxy distribution.

Distances were calculated assuming the absorption systems to have no
peculiar velocities with respect to the center of the voids. Although
this assumption might be realistic for gas inside voids, it might not
be the case for gas residing in denser environments, where gas outflows
from galaxies might dominate. However, some studies have suggested that
the bulk of \lya~forest lines have little velocity offset with respect
to galaxies \citep[e.g.,][]{theuns02,wilman07}. As an example, a
velocity difference of $\sim 200$ \kms~at redshift $z_{\rm abs}\lesssim
0.1$ would give an apparent distance shift of the order of $\sim 2$
\mpc, which is somewhat higher than, but comparable to the systematic
error of the void center determination (given that voids regions are
not perfectly spherical as assumed here). Note that the uncertainty in
the void center is smaller than the uncertainty of a single galaxy
because the void is defined by an average over many galaxies. As
previously mentioned, such an uncertainty should not artificially
create a false signal but rather should dilute any real difference.

Although the $X$ and $D$ coordinates are not independent, we decided to
show our results using both. This has the advantage of testing the
consistency of our results using two slightly differently motivated
definitions. $X$ is a scaled coordinate, good for stacking voids of
different radius. It is also good for comparisons with some of the P12
results. $D$ gives a direct measure of the actual distances involved,
while still using $R_{\rm void}$. For convenience, results associated
with the $X$ definition will be shown normally in the text while
results associated with the $D$ definition will be shown in
parenthesis: $X$ ($D$) format.

Figure \ref{fig:X_D_dist} shows the histogram of absorption systems as
a function of $X$ and $D$ (left and right panel respectively). In order
to show the effects of the geometry of the survey, the random
expectations are also shown (shaded distributions). To generate the
random samples, we placed $1000$ random absorption systems per real
one, uniformly between $z_{\rm lim} < z_{\rm abs} \le 0.102$ for each
sightline, where $z_{\rm lim}$ corresponds to the maximum of $0.01$ and
the minimum observed redshift for a \lya~in that sightline. In this
calculation we have masked out spectral regions over a velocity window
of $\pm200$ \kms~around the position where strong Galactic absorption
could have been detected (namely: C~{\sc i}, C~{\sc ii}, N~{\sc v},
O~{\sc i}, Si~{\sc ii}, P~{\sc iii}, S~{\sc i}, S~{\sc ii}, Fe~{\sc ii}
and Ni~{\sc ii}) {\it before} the random redshifts are assigned. A
total of $106\,000$ random absorbers were generated. We observe a
relative excess of absorption systems compared to the random
expectation between $X \simeq 0.9$--$1.3$ and/or $D \simeq -2$--$4$
\mpc. Assuming Poisson uncertainty, there were $61 \pm 8$ ($65 \pm
8$)\footnote{Results regarding distances from the center of voids are
  presented in a $X$ ($D$) format (see \S
  \ref{gasaroundvoids}). Reference to this footnote will be omitted
  hereafter.} observed, while $\approx 38.3 \pm 0.2$ ($42.4 \pm
0.2$)\fn~were expected from the random distribution. This corresponds
to an $\approx 3\sigma$ excess. Similarly, there is a significant
($\approx 3\sigma$) deficit of absorption systems at $X \gtrsim 1.3$
and/or $D \gtrsim 4$ \mpc, for which $17 \pm 4$ ($19 \pm 4$)\fn~systems
were observed compared to the $33.9 \pm 0.2$ ($34.1 \pm
0.2$)\fn~randomly expected. We also checked that such an excess and
deficit did not appear by chance in $1000$ realizations, consistent
with the $<0.1\%$ probability of occurrence. No significant difference
is found for systems at $X \lesssim 0.9$ and/or $D \lesssim -2$ \mpc,
for which the $28 \pm 5$ ($22 \pm 5$)\fn~found are consistent with the
random expectation of $33.4 \pm 0.2$ ($29.0 \pm 0.2$)\fn. The
Kolmogorov-Smirnov (KS) test between the full unbinned samples gives a
$\approx 0.3\%$ ($0.5\%$)\fn~probability that both the random and the
real data come from the same parent distribution. We checked that no
single sightline dominates the signal by removing each individual one
and repeating the previous calculation. We also checked that masking
out the spectral regions associated to possible Galactic absorption
does not have an impact on our results as the same numbers (within the
errors) are recovered when these regions are not excluded. These
results hint at a well defined gas structure around voids, possibly
analogous to that seen in galaxies. The current data are not sufficient
to confirm (at a high confidence level) the reality of the apparent
two-peaked shape seen in the real distributions
however.\begin{figure*}
  \includegraphics[width=160mm]{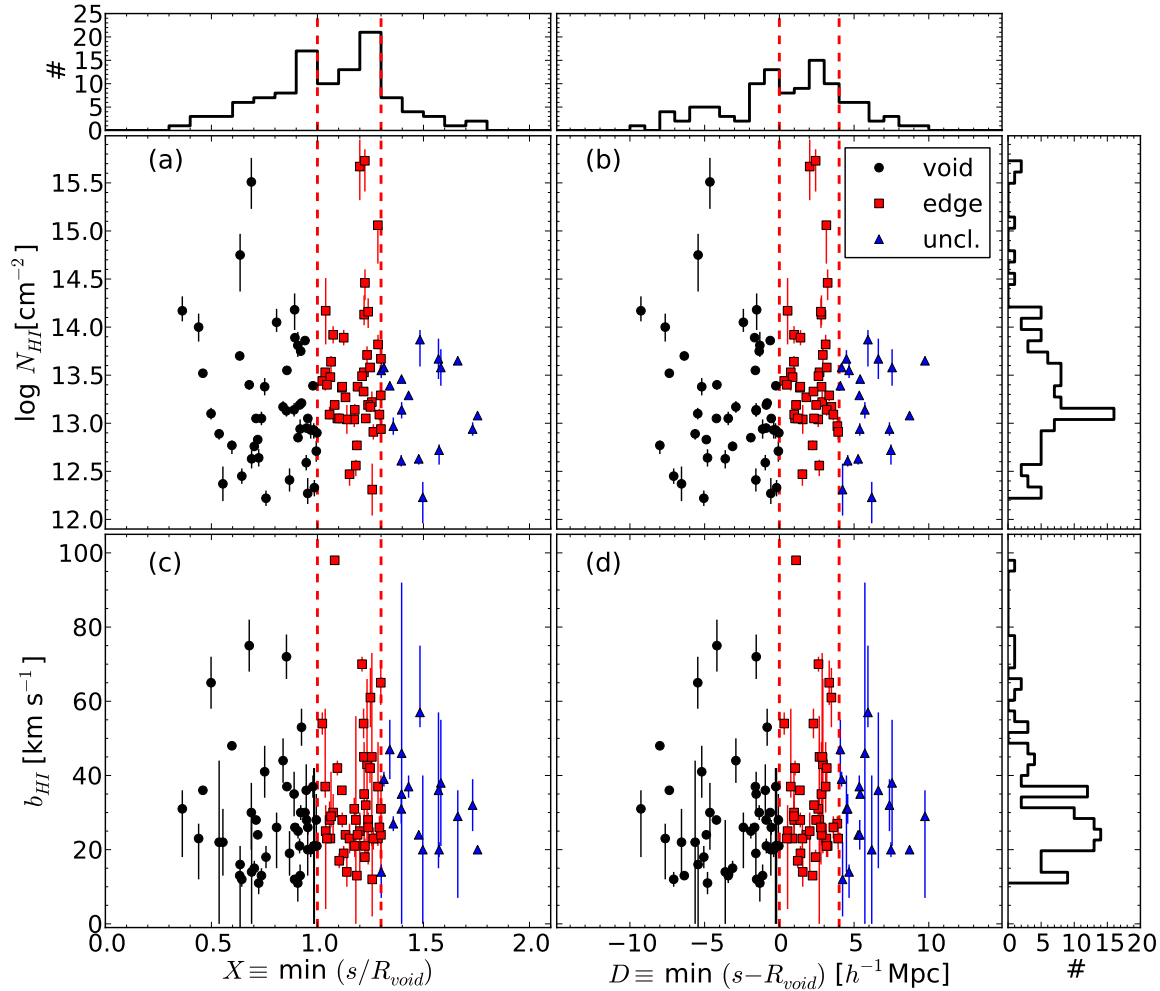}
  \caption{Panels (a) and (b) show the distribution of column densities
    of \hi~as a function of $X$ and $D$ respectively. Panels (c) and
    (d) show the distribution of Doppler parameters as a function of
    $X$ and $D$ respectively. Our LSS samples are shown by different
    color/symbols: void (black circles), void-edge (red squares) and
    unclassified (blue triangles). Histograms are also shown around the
    main panels. Vertical red dashed lines show the limits of our LSS
    definitions (see \S \ref{definition}).}\label{fig:2}

\end{figure*}

\begin{table*}
  \centering
  \begin{minipage}{140mm}
    \caption{General properties of our LSS samples\tablenotemark{a}.}\label{tab:2}
\begin{tabular}{lcccccc}
  \hline
  Sample                &\multicolumn{2}{c}{$\log(N_{HI}[\rm{cm}^{-2}])$} & \multicolumn{2}{c}{$b_{HI}$ [\kms]}  \\
                  & mean                                 & median   & mean & median \\
  \hline
  Void            & 13.21 $\pm$ 0.67 (13.21 $\pm$ 0.67)  & 13.05 (13.05)  &28 $\pm$ 15 (28 $\pm$ 15)&25  (25)\\
  Edge            & 13.50 $\pm$ 0.70 (13.52 $\pm$ 0.69)  & 13.38 (13.38)  &33 $\pm$ 17 (34 $\pm$ 17)&28  (28)\\
  Unclassified            & 13.20 $\pm$ 0.45 (13.17 $\pm$ 0.48)  & 13.36 (13.36)  &33 $\pm$ 11 (31 $\pm$ 11)&32  (31)\\
      
  \hline
\end{tabular}
\tablenotetext{a}{Results are presented in a $X$ ($D$) format (see \S
  \ref{gasaroundvoids}).}
\end{minipage}
\end{table*}

\subsection{Definition of large scale structure in absorption}\label{definition}
We define three LSS samples observed in absorption:

\begin{itemize}
\item {\it Void absorbers:} those absorption systems with $X<1$ and/or
  $D<0$ \mpc. A total of $45$ ($45$)\fn~void absorbers were found.

\item {\it Void-edge absorbers:} those absorption systems with $1 \le X
  < 1.3$ and/or $0 \le D<4$ \mpc. A total of $44$ ($42$)\fn~void-edge
  absorbers were found.

\item {\it Unclassified absorbers:} those absorption systems with $X
  \ge 1.3$ and/or $D \ge 4$ \mpc. A total of $17$ ($19$)\fn~unclassified
  absorbers were found.
\end{itemize}

\noindent To be consistent with the galaxy-void definition, we use $X =
1$ and/or $D=0$ \mpc~as the limits between void and void-edge
absorbers. The division between void-edge and unclassified absorbers
was chosen to match the transition between the overdensity to
underdensity of observed absorbers compared to the random expectation
at $X>1$ and/or $D>0$ \mpc~(see Figure \ref{fig:X_D_dist}).

We have assumed here that the center of galaxy voids will roughly
correspond to the center of gas voids, however that does not
necessarily imply that gas voids and galaxy voids have the same
geometry. In fact, as we do not find a significant underdensity in the
number of void absorbers with respect to the random expectation, it is
not clear that such voids are actually present within the \lya~forest
population. Of course, the fact that we do not detect this
under-density, does not imply that the gas voids are not there. A
better way to look at these definitions is by considering void
absorbers as those found in galaxy under-densities (galaxy voids) and
void-edge absorbers as those found in regions with a typical density of
galaxies. We do not have a clear picture of what the unclassified
absorbers correspond to. Unclassified absorbers are those lying at the
largest distances from the {\it cataloged} voids, but this does not
necessarily imply that they are associated with the highest density
environments only. In fact, there could be high density regions also
located close to void-edges, at the intersection of the cosmic web
filaments. Given that voids of radius $\lesssim 10$ \mpc~are not
present in the current catalog it is also likely that some of the
unclassified absorbers are associated with low density
environments. Therefore, one interpretation of unclassified-absorbers
could be as being a mixture of all kind of environments, including
voids, void-edges and high density regions.

We checked the robustness of these definitions by looking at the number
of voids and void-edges which can be associated with a given
absorber. In other words, for a given absorption system, we counted how
many voids or void-edges could have been associated with it by taking
simply $X \equiv s/R_{\rm void}$ or $D \equiv s - R_{\rm void}$ (in
contrast to having taken the minimum values). Out of the $45$ void
absorbers, $41$ are associated with only one void and $4$ are
associated with $2$ voids, independently of the definition used (either
$X$ or $D$). Likewise, out of the $44$ ($42$)\fn~void-edge absorbers,
$31$ ($28$)\fn~are associated with just one void-edge, $12$
($13$)\fn~are associated with two void-edges and $1$ ($1$)\fn~is
associated with three void-edges. This last system is located at
$X=1.04$ ($D=0.55$ \mpc) and has \nhi~$=10^{14.17 \pm 0.35}$ \cm~and
\bhi~$=25^{+21}_{-7}$ \kms~at a redshift of $z_{\rm abs}=0.01533$. From
these values the system does not seem to be particularly
peculiar. Finding an association with more than two void-edges is not
surprising as long as the filling factor of voids is not
small\footnote{For reference, voids found by P12 have a filling factor
  of $62\%$.}. Void absorbers have on average $1.1 \pm 0.3$ voids
associated with them, with a median of $1$. Void-edge absorbers have in
average $1.3 \pm 0.5$ ($1.4 \pm 0.5$)\fn~void-edges associated with
them, with a median of $1$($1$)\fn. These values give a median
one-to-one association. Therefore, we conclude then that the LSS
definitions used here are robust.

\subsection{Properties of absorption systems in different large scale structure regions}

Figure \ref{fig:2} shows the distribution of column densities and
Doppler parameters as a function of both $X$ and $D$. At first sight,
no correlation is seen between \nhi~or \bhi~and distance to the center
of voids. Table \ref{tab:2} gives the mean and median values of
$\log(N_{HI}[\rm{cm}^{-2}])$ and \bhi~ for our void, void-edge and
unclassified absorption systems. These results show consistency within
$1\sigma$ between the three LSS samples.

\begin{table*}
  \centering
  \begin{minipage}{140mm}
    \caption{Kolmogorov-Smirnov (KS) test probabilities between
      different samples\tablenotemark{a}.}\label{tab:3}
\begin{tabular}{lcccccc}
  \hline
                             & void/edge   &void/uncl.   &  edge/uncl.   &void/not-void&edge/not-edge & uncl./not-uncl. \\
  \hline
  KS-Prob($\log N_{\rm HI}$) & 2\% (0.7\%)       & 74\% (66\%)&  56\% (24\%) & 4\% (4\%)   & 3\% (0.6\%)    & 64\% (54\%) \\ 
  KS-Prob($b_{\rm HI}$)       & 8\% (6\%)        & 18\% (17\%)&  71\% (75\%) & 7\% (7\%)   & 20\% (14\%)    & 32\% (32\%) \\ 
  \hline

\end{tabular}
\tablenotetext{a}{Results are presented in a $X$ ($D$) format (see \S
  \ref{gasaroundvoids}).}
\end{minipage}
\end{table*}

A closer look at the problem can be taken by investigating the possible
differences in the full \nhi~and \bhi~distributions of void, void-edge
and unclassified absorbers.

\begin{figure}
  \includegraphics[width=90mm]{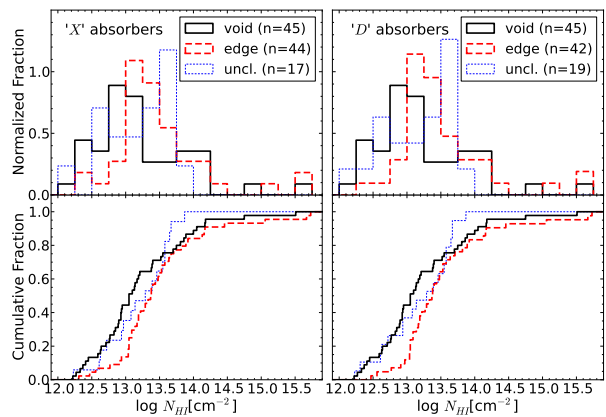}
  \caption{\hi~column density distribution for the three different LSS
    defined in this work (see \S \ref{definition}): void absorbers
    (solid-black lines), void-edge absorbers (red-dashed lines) and
    unclassified absorbers (blue-dotted lines). Top panels show the
    normalized distribution using arbitrary binning of $0.5$
    dex. Bottom panels show the cumulative distributions for the
    unbinned samples. Left and right panels correspond to absorbers
    defined using $X$ and $D$ coordinates respectively.}\label{fig:3}

\end{figure}

\subsubsection{Column density distributions}\label{cdd}
Figure \ref{fig:3} shows the distribution of column density for the
three different LSS defined above (see \S \ref{definition}). The top
panels show the normalized fraction of systems as a function of
\nhi~(arbitrary binning), whilst the bottom panels show the cumulative
distributions (unbinned). We see from the top panels that this
distribution seems to peak systematically at higher \nhi~from void to
void-edge and from void-edge to unclassified absorbers. We also observe
a suggestion of a relative excess of weak systems ($N_{\rm HI} \lesssim
10^{13}$ \cm) in voids compared to those found in void-edges. This can
also be seen directly in Figure \ref{fig:2} (see panels (a) and
(b)). The KS test gives a probability $P^{\log N}_{\rm void/edge}
\approx 2\%$ ($0.7\%$)\fn~that void and void-edge absorbers come from
the same parent distribution. This implies a $>2\sigma$ difference
between these samples. No significant difference is found between voids
or void-edges with unclassified absorbers, for which the KS test gives
probabilities of $P^{\log N}_{\rm void/uncl.} \approx 74\%$
($66\%$)\fn~and $P^{\log N}_{\rm edge/uncl.} \approx 56\%$
($24\%$)\fn~respectively. These results can be understood by looking at
the bottom panels of Figure \ref{fig:3}, as we see that the maximum
difference between the void and void-edge absorbers distributions is at
$N_{\rm HI} \lesssim 10^{13.8}$ \cm. On the other hand, no big
differences are observed at $N_{\rm HI} \gtrsim 10^{13.8}$ \cm. In
fact, by considering just the systems at $N_{\rm HI} < 10^{13.8}$ \cm,
the significance of the difference between void and void-edge absorbers
is increased, with $P^{\log N}_{\rm void/edge} \approx 0.9\%$
($0.2\%$)\fn. Likewise, at $N_{\rm HI} \ge 10^{13.8}$ \cm, void and
void-edge absorber distributions agree at the $\approx 86\%$
($86\%$)\fn~confidence level. We note however that there were $\le 10$
systems per sample for this last comparison and therefore, it is likely
to be strongly affected by low number statistics.

We also investigated possible differences between void, void-edge and
unclassified absorbers and their complements (i.e., all the systems
that were not classified as these: not-void, not-void-edge,
not-unclassified). Not-voids correspond to the combination of void-edge
and unclassified absorbers and so on. The KS gives probabilities of
$P^{\log N}_{\rm void/not-void} \approx 4\%$ ($4\%$)\fn, $P^{\log
  N}_{\rm edge/not-edge} \approx 3\%$ ($0.6\%$)\fn~implying that void
and void-edge absorbers are somewhat inconsistent with their
complements. On the other hand, the distribution of unclassified
absorbers is consistent with the distribution of their complements with
a KS probability of $P^{\log N}_{\rm uncl./not-uncl.} \approx 64\%$
($54\%$)\fn. These results are summarized in Table \ref{tab:3}.

\begin{figure}
  \includegraphics[width=90mm]{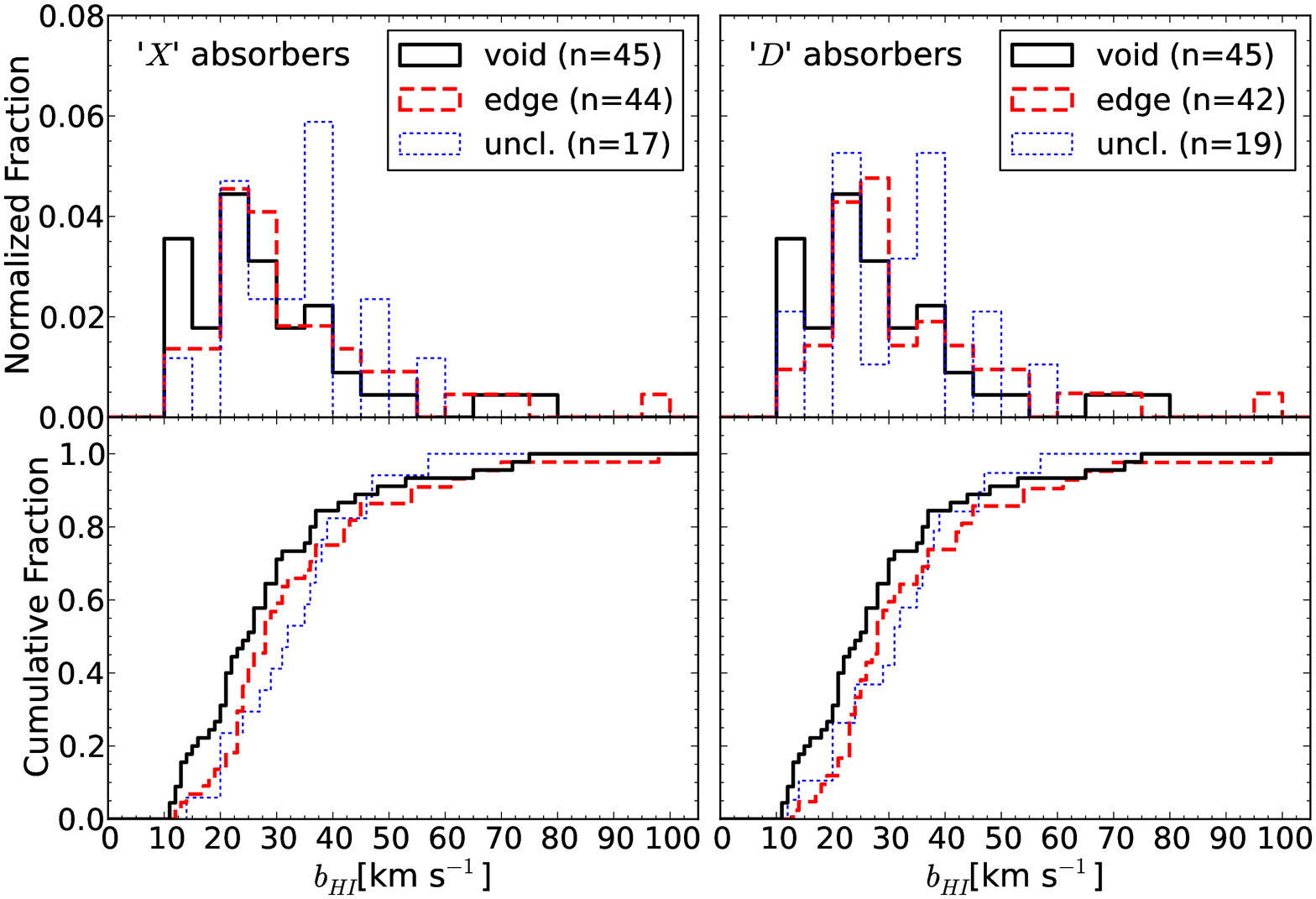}
  \caption{\hi~Doppler parameter distribution for the three different
    LSS defined in this work (see \S \ref{definition}): void absorbers
    (black-solid lines), void-edge absorbers (red-dashed lines) and
    unclassified absorbers (blue-dotted lines). Top panels show the
    normalized distribution using arbitrary binning of $5$ \kms. Bottom
    panels show the cumulative distributions for the unbinned
    samples. Left and right panels correspond to absorbers defined
    using $X$ and $D$ coordinates respectively.}\label{fig:4}

\end{figure}

\subsubsection{Doppler parameter distributions}\label{dpd}

Figure \ref{fig:4} shows the distribution of Doppler parameter for the
three different LSS defined above (see \S \ref{definition}). The top
panels show the normalized fraction of systems as a function of
\bhi~(arbitrary binning), whilst bottom panels show the cumulative
distributions (unbinned). This figure suggests a relative excess of
low-\bhi~systems ($b_{\rm HI} \lesssim 20$ \kms) in voids compared to
those from void-edge and unclassified samples. A relative excess of
unclassified absorbers compared to that of voids or void-edges at
high-\bhi~values ($b_{\rm HI} \gtrsim 35$ \kms) is also suggested by
the figure. The KS test gives a probability $P^{b}_{\rm void/edge}
\approx 8\%$ ($6\%$)\fn~that void and void-edge absorbers come from the
same parent distribution. This implies no detected difference between
void and void-edge absorbers. Likewise, no significant difference is
found between voids or void-edges with unclassified absorbers, for
which the KS test gives probabilities of $P^{b}_{\rm void/uncl.}
\approx 18\%$ ($17\%$)\fn~and $P^{b}_{\rm edge/uncl.} \approx 71\%$
($75\%$)\fn~respectively.

As before, we also investigated possible difference between LSS and
their complements. In this case, neither void, void-edge or
unclassified absorbers are significantly different than their
complements with KS probabilities of $P^{b}_{\rm void/not-void} \approx
7\%$ ($7\%$)\fn, $P^{b}_{\rm edge/not-edge} \approx 20\%$
($14\%$)\fn~and $P^{b}_{\rm uncl./not-uncl.} \approx 32\%$
($32\%$)\fn. These results are also summarized in Table \ref{tab:3}.

\subsection{Check for systematic effects}
Given that the differences between void and void-edge samples are still
at $<3\sigma$ of confidence level, we have investigated possible biases
or systematic effects that could be present in our data analysis. In
particular we have investigated (1) possible differences in our
subsample with respect to the whole DS08 sample, (2) the effect of the
different characterization methods used by DS08 to infer the gas
properties, and (3) whether uniformity across our redshift range is
present in our observables. A complete discussion is presented in
Appendix \ref{systematic}. From that analysis we concluded that no
important biases affect our results.

\section{Comparison with Simulations}\label{gimic}
In this section, we investigate whether current cosmological
hydrodynamical simulations can reproduce our observational results
presented in \S \ref{analysis}. For this comparison we use the {\it
  Galaxies-Intergalactic Medium Interaction Calculation}
\citep[\gimic,][]{crain09}. Using initial conditions drawn from the
      {\it Millennium} simulation \citep{springel05}, \gimic~follows
      the evolution of baryonic gas within five, roughly spherical
      regions (radius between $18-25$ $h^{-1}$ Mpc\footnote{Note that
        \gimic~adopted a $H_0=100$ $h$ \kms Mpc$^{-1}$,$h=0.73$,
        $\Omega_{\rm m}=0.25$, $\Omega_{\rm \Lambda}=0.75$,
        $\sigma_8=0.9$, $k=0$ cosmology. These parameters are slightly
        different than the ones used in P12.}) down to $z=0$ at a
      resolution of $m_{\rm gas} \approx 10^7$ $h^{-1}$ M$_{\sun}$. The
      regions were chosen to have densities deviating by
      $(-2,-1,0,+1,+2)\sigma$ from the cosmic mean at $z=1.5$, where
      $\sigma$ is the rms mass fluctuation. The $+2\sigma$ region was
      additionally required to be centered on a rich cluster
      halo. Similarly although not imposed, the $-2\sigma$ region is
      approximately centered on a sparse void. The rest of the
      Millennium simulation volume is re-simulated using only the dark
      matter particles at much lower resolution to account for the
      tidal forces. This approach gives \gimic~the advantage of probing
      a wide range of environments and cosmological features with a
      comparatively low computational expense.

\gimic~includes (i) a recipe for star formation designed to enforce a
local Kennicutt-Schmidt law \citep{schaye08}; (ii) stellar evolution
and the associated delayed release of $11$ chemical elements
\citep{wiersma09b}; (iii) the contribution of metals to the cooling of
gas in the presence of an imposed UV background \citep{wiersma09a}; and
(iv) galactic winds that pollute the IGM with metals and can quench
star formation in low-mass halos \citep{dallavecchia08}. Note that
\gimic~does not include feedback processes associated with AGN. For
further details about \gimic~we refer the reader to \citet{crain09}.

\subsection{Simulated \hi~absorbers sample}\label{gimic:simulated}
In order to obtain the properties of the simulated \hi~absorption
systems, we placed $1000$ parallel sightlines within a cube of $20$
$h^{-1}$Mpc on a side centered in each individual \gimic~region at
$z=0$ ($5000$ sightlines in total). We have excluded the rest of the
volume to avoid any possible edge effects. This roughly corresponds to
$2.5$ sightlines per square $h^{-1}$Mpc. Given this density, some
sightlines could be tracing the same local LSS and therefore these are
not fully independent. We consider this approach to offer a good
compromise of having a large enough number of sightlines while not
oversampling the limited \gimic~volumes.

We used the program {\sc specwizard}\footnote{Written by Joop Schaye,
  Craig M. Booth and Tom Theuns.} to generate synthetic normalized
spectra associated to our sightlines using the method described by
\citet{theuns98}. {\sc specwizard} calculates the optical depth as a
function of velocity along the line-of-sight, which is then converted
to flux transmission as a function of wavelength for a given
transition. We only used \hi~in this calculation. The spectra were
convolved with an instrumental spread function (Gaussian) with FWHM of
$6.6$ \kms~to match the resolution of the STIS/HST
spectrograph\footnote{Note that the majority of the \lya~used in this
  work were observed with STIS/HST rather than FUSE.}. In order to
mimic the continuum fitting process in real spectra, we set the
continuum level of each mock {\it noiseless} spectrum at the largest
flux value after the convolution with the instrumental profile. Given
that the lines are sparse at $z=0$, there were almost always regions
with no absorption and this last correction was almost negligible.

We used 3 different signal-to-noise ratios in order to represent our
QSO sample. Out of the total of $1000$ per \gimic~region, $727$
sightlines were modeled with $S/N=9$ per pixel, $182$ with $S/N=23$ and
$91$ $S/N=2$. These numbers keep the proportion between the different
$S/N$ values as it is in the observed sample (see last column in Table
\ref{tab:1})\footnote{Note that we have divided the mean $S/N$ per
  two-pixel resolution element by $\sqrt{2}$ to have an estimation per
  pixel.}.

We fit Voigt profiles to the synthetic spectra automatically using {\sc
  vpfit}\footnote{Written by R.F. Carswell and J. K. Webb (see
  http://www.ast.cam.ac.uk/$\sim$rfc/vpfit.html).}, following the
algorithm described by \citet{crighton10}. First, an initial guess of
several absorption lines is generated in each spectrum to minimize
$\chi^2_{\rm reduced}$. If the $\chi^2_{\rm reduced}$ is greater than a
given threshold of $1.1$, another absorption component is added at the
pixel of largest deviation and $\chi^2_{\rm reduced}$ is
re-minimized. Absorption components are removed if both
\nhi$<10^{14.3}$ \cm~and \bhi$<0.4$ \kms. This iteration continues
until $\chi^2_{\rm reduced} \le 1.1$. Then, the Voigt fits are
stored. We only kept absorption lines where the values of $\log N_{\rm
  HI}$ and \bhi~are at least $5$ times their uncertainties as quoted by
{\sc vpfit}.

\begin{figure*}
  \includegraphics[width=160mm]{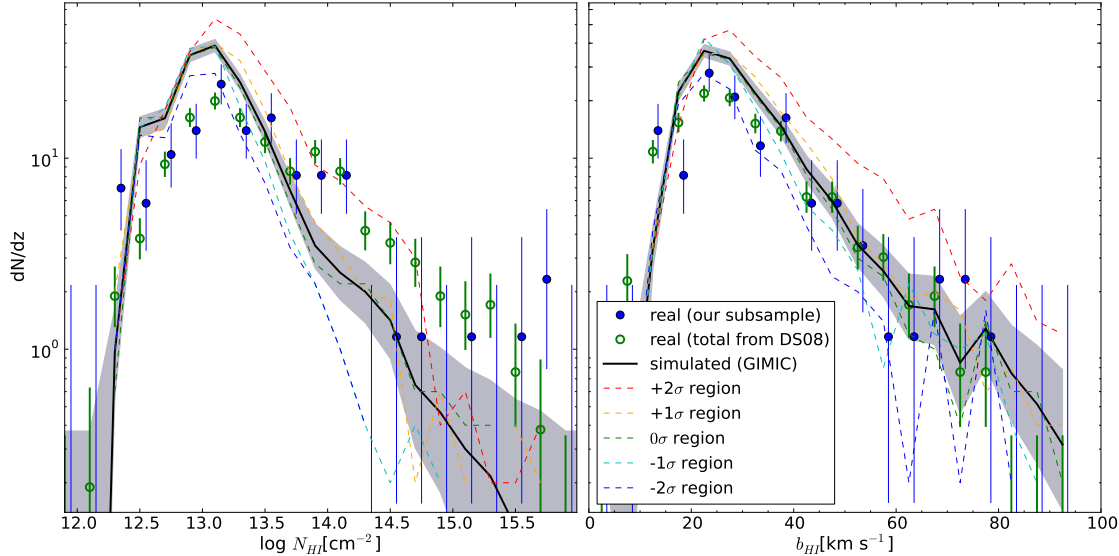}
  \caption{Redshift number density of \hi~lines as a function of both
    column density (left panel) and Doppler parameter (right panel)
    using abritrary binning of $\Delta \log N_{\rm HI} = 0.2$ dex and
    $\Delta b_{\rm HI} = 5$ \kms~respectively. Both results have not
    been corrected for incompleteness. Green open-circles correspond to
    real data from the total sample of DS08 ($657$ systems) while blue
    filled-circles (slightly offset in the $x$-axes for clarity)
    correspond to the subsample used in this study ($106$ systems). The
    black line corresponds to the volume-weighted result from the
    combination of the five \gimic~regions where the shaded region
    correspond to the $\pm 1\sigma$ uncertainty. Dashed lines show the
    results from each individual \gimic~region. Error bars correspond
    to the Poissonian uncertainty from the analytical approximation
    $\sigma^+_n\approx \sqrt{n + \frac{3}{4}}+1$ and $\sigma^-_n\approx
    \sqrt{n - \frac{1}{4}}$ \citep{gehrels86}.} \label{fig:7}

\end{figure*}

The fraction of hydrogen in the form of \hi~within \gimic~is obtained
from {\sc cloudy} \citep{ferland98} after assuming an ionization
background from \citet{haardt96} that yields a photo-ionization rate
$\Gamma = 8.59 \ 10^{-14}$ s$^{-1}$. This ionization background is not
well constrained at $z\approx0$, so we use a post processing correction
to account for this uncertainty. In the optically thin regime
$\Gamma^{\rm thin} \propto \frac{1}{\tau}$, where $\tau$ is the optical
depth \citep{gunn65}. Then, scaling the optical depth values is
equivalent to scaling the ionization background
\citep[e.g.,][]{theuns98b,dave99}. First, we combined the five
\gimic~regions using different volume weights namely:
$(1/12,1/6,1/2,1/6,1/12)$ for the $(-2,-1,0,+1,+2)\sigma$ regions
respectively \citep[see Appendix 2 in][for a justification of these
  weights]{crain09}. Then, we searched for a constant value to scale
all the original optical depth values such that the mean flux of the
combined sample is equal to the observed mean flux of \lya~absorption
at low redshift. A second possibility is to scale the optical depth
values in order to match the redshift number density of \hi~lines in
some column density range, $dN/dz$, instead of the mean flux. Ideally
by matching one observable the second would be also matched.

Extrapolating the double power-law fit result from \citet{kirkman07} to
$z=0$ (see their equation $6$), the observed mean flux is $\langle F
\rangle = 0.987$ with a typical statistical uncertainty of
$\sigma_{\langle F \rangle} \sim 0.003$. In order to match this number
in the simulation a scale of $1.16$ is required in the original optical
depth values ($0.86$ in $\Gamma$).  From this correction, the redshift
number density of lines in the range $10^{13.2} \le N_{\rm HI} \le
10^{14}$ \cm~is found to be $dN/dz \approx 50$. For reference,
\citet{lehner07} and DS08 found $dN/dz \sim 50 - 90 $ over the same
column density range. We have repeated the experiment for consistency,
using $\langle F \rangle = \{0.984,0.990\}$ which are within $\pm
1\sigma_{\langle F \rangle}$ of the extrapolated value. To match these,
scales of $1.50$ and $0.84$ are required in the original optical depths
values respectively ($0.67$ and $1.19$ in $\Gamma$). From those mean
fluxes we found $dN/dz \approx \{70, 35\}$ respectively along the same
column density range. Therefore, a value of $\langle F \rangle = 0.990$
underpredicts the number of \hi~lines. On the other hand, values of
$\langle F \rangle = 0.987$ and $0.984$ are in good agreement with
observations. In the following analysis we use $\langle F \rangle =
0.987$ unless otherwise stated.

\subsection{Comparison between simulated and observed \hi~properties}\label{gimic:properties}

Figure \ref{fig:7} shows the redshift number density of \hi~lines (not
corrected for incompleteness) as a function of both column density
(left panel) and Doppler parameter (right panel). Data from the
simulation are shown by the black line (volume-weighted result) and
each individual \gimic~region is shown separately by the dashed
lines. For comparison, data from observations are also shown. Green
open-circles correspond to the total sample from DS08 ($657$ systems)
while blue filled-circles correspond to the subsample used in this
study ($106$ systems that intersect the SDSS volume). There is not
perfect agreement between simulated and real data. We see an excess
(lack) of systems with \nhi~$\lesssim 10^{13.5}$ \cm~(\nhi~$\gtrsim
10^{14}$ \cm) in the simulation compared to observations while Doppler
parameters are in closer agreement, although there is still a
difference at low \bhi.

\begin{figure*}
  \includegraphics[width=160mm]{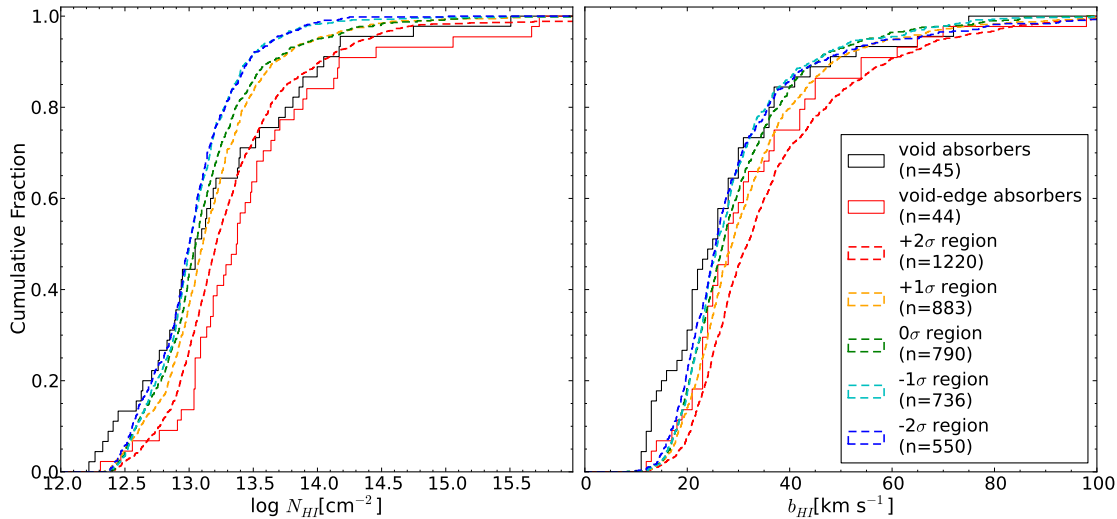}
  \caption{Column density (left panel) and Doppler parameter (right
    panel) cumulative distributions for \hi. Void absorbers are shown
    by solid-black lines while void-edge absorbers are shown by
    solid-red lines. Dashed lines show the result from each individual
    \gimic~region. For simplicity we only show LSS definitions based on
    $X$.}\label{fig:8}

\end{figure*}

Assuming that the column density distribution can be modeled as a
power-law, the position of the turnover at the low \nhi~end give us an
estimation of the completeness level of detection in the sample. As the
turnover appears to be around \nhi~$\approx 10^{13}$ \cm~in both
simulated and real data (by design) we do not, in principle, attribute
the discrepancy in the column-density distributions to a wrong choice
of the simulated $S/N$. Raising the mean flux to a greater value than
$\langle F \rangle = 0.987$ (less absorption) does not help as the
$dN/dz$ in the range $10^{13.2} \le N_{\rm HI} \le 10^{14}$ \cm~will
then be smaller than the observational result (see \S
\ref{gimic:simulated}). We attempted to get a better match by using a
mean flux of $\langle F \rangle = 0.984$ (more absorption), motivated
to produce a better agreement at higher column densities. In order to
agree at low column densities, we had to degrade the sample $S/N$ to be
composed of $\sim400$, $\sim100$ and $\sim500$ sightlines at
signal-to-noise ratios of $9$, $23$ and $2$ respectively. It is
implausible that half of the observed redshift path has such poor
quality.

Another possibility to explain the discrepancy could be the fact that
weak systems in observations were preferentially characterized with the
AOD method, whereas here we have only used Voigt profile fitting. In
order to test this hypothesis, we have merged closely separated systems
(within $150$ \kms) whose summed column density is less than
$10^{13.5}$ \cm. Using these constraints, $43$ out of $4179$ systems
were merged ($\approx 1\%$). Such a small fraction does not have an
appreciable effect on the discrepancy. As an extreme case, we have
repeated the experiment merging {\it all} systems within $300$
\kms~independently of their column densities. From this, $555$ out of
$4179$ systems were merged ($\approx 13\%$) but still it was not enough
to fully correct the discrepancy. Given that the discrepancy is not
explained by a systematic effect from different line characterization
methods, we chose to keep our original simulated sample in the
following analysis without merging any systems.

There is a reported systematic effect by which column densities
inferred from a single \lya~line are typically (with large scatter)
underestimated with respect to the curve-of-growth (COG)
solution. Similarly \bhi~are typically overestimated
(\citealt{shull00}; \citealt{danforth06}; see also Appendix
\ref{systematic:method} for discussion on how this may affect our
observational results). This effect is only appreciable for
\nhi~$\gtrsim 10^{14}$ \cm~and is bigger for saturated lines. Given
that our simulated sample was constructed to reproduce the observed
sample, this effect could be present. If so, it would in principle help
to reduce the discrepancy at the high column density end. From figure 3
of \citet{danforth06} we have inferred a correction for systems with
\nhi~$\ge 10^{13.5}$ \cm~of,

\begin{equation}
\log N_{\rm HI}^{\rm corr} = \frac{\log N_{\rm HI}^{\rm obs} - 8.37}{1
  - 0.62} \ \rm{,}
\label{eq:logNcorrection}
\end{equation}

\noindent where $N_{\rm HI}^{\rm corr}$ and $N_{\rm HI}^{\rm obs}$ are
the corrected and observed \nhi~values respectively. From this
correction we found an increase in the number of systems at
\nhi~$\gtrsim 10^{14.5}$ \cm~up to values consistent with
observations. This however does not help with the discrepancy at lower
column densities.

At this point, it is difficult to reconcile the simulation result with
the real data using only a single effect. We note that the discrepancy
is a factor of $\sim 2$ only, so it could be in principle explained
through a combination of several observational effects.  Also note that
the number of observed lines at higher column densities is still small
and it could be affected by low number statistics. The lack of systems
with very low \nhi~and \bhi~values can be explained by our selection of
the highest signal-to-noise being $S/N=23$ while in real data there
could be regions with higher values. It is not the aim of this section
to have a perfect match between simulations and observations but rather
examine the qualitative differences between simulated regions of
different densities. Thus hereafter, we will use the results from the
simulation in its original form (as shown in Figure \ref{fig:7}), i.e.,
without any of the aforementioned corrections.

\subsection{Simulated \hi~absorbers properties in different LSS regions}

Given that \gimic~does not provide enough volume to perform a
completely analogous search for voids (each region is $\sim20$
$h^{-1}$Mpc of radius), we use them only as crude guides to compare our
results with. We could consider the $-2\sigma$ region as representative
of void regions as it is actually centered in one. Naively, we could
consider the $0\sigma$ regions as representative of void-edge regions,
as it is there where the mean cosmological density is reached. A direct
association for the $+1\sigma$ and $+2\sigma$ is not so simple though,
as they would be associated to some portions of the void-edge regions
too. It seems more reasonable to use the \gimic~spheres as
representative of different density environments then, where
$-2\sigma$/$+2\sigma$ correspond to extremely under/over-dense regions
and so on. For reference, the whole $(-2,-1,0,+1,+2)\sigma$
\gimic~regions correspond to densities of $\frac{\rho}{\langle \rho
  \rangle}\approx (0.4,0.6,0.9,1.2,1.8)$\footnote{Given that we are
  using cubic sub-volumes centered in these spheres, these cubes should
  have higher density differences between them.}  at $z=0$
respectively, where $ \langle \rho \rangle$ is the mean density of the
universe \citep[see Figure A1 from][]{crain09}.

Figure \ref{fig:8} shows the cumulative distributions of \nhi~(left
panel) and \bhi~(right panel). Results from each of the individual
\gimic~region are shown by dashed lines. Void and void-edge absorbers
are shown by solid black and red lines respectively. For simplicity we
show only LSS definition based on $X$. Cumulative distributions between
real and simulated data do not agree perfectly. However, in both real
and simulated data, there is an offset between column densities and
Doppler parameters found in different environments. Low density
environments have smaller values for both \nhi~and \bhi~than higher
density ones (and viceversa). This trend still holds when using a
$S/N=23$ per pixel for the $5000$ sightlines.

\begin{table}
  \centering
  \begin{minipage}{80mm}
    \caption{Kolmogorov-Smirnov (KS) test probabilities between
      different \gimic~regions\tablenotemark{a}.}\label{tab:4}
    \begin{tabular}{@{}c|rrrrr@{}}
  \hline
   \backslashbox{$N_{\rm HI}$}{$b_{\rm HI}$} 
             & $-2\sigma$   &$-1\sigma$   &  $\ 0\sigma$  &$+1\sigma$ & $+2\sigma$     \\
  \hline                                                                               \\
  $-2\sigma$ &   \nodata        &   $30\%$     &   $2\%$     & $2\%$     &  $\ll 1\%$  \\
  $-1\sigma$ &  $77\%$     &     \nodata        &   $6\%$     & $\ll 1\%$ &  $\ll 1\%$ \\
  $0\sigma$  &  $11 \%$    &   $0.9\%$    &     \nodata       & $0.7\%$   &  $\ll 1\%$ \\
  $+1\sigma$ &  $\ll 1\%$  &   $\ll 1\%$  &   $0.4\%$   &     \nodata     &  $\ll 1\%$ \\
  $+2\sigma$ &  $\ll 1\%$  &   $\ll 1\%$  &   $\ll 1\%$ & $\ll 1\%$  &    \nodata      \\
  \hline
  \end{tabular}
    \tablenotetext{a}{KS test probabilities for $\log N_{\rm HI}$ and
      \bhi~distributions are shown in the left-bottom and right-upper
      sides respectively. $\ll 1\%$ corresponds to values $<10^{-4}
      \%$.}
\end{minipage}
\end{table}

The KS test gives a significant difference between the $+2\sigma$,
$+1\sigma$ and $+0\sigma$ regions, and any other \gimic~region at the
$\gg 99\%$, $\gtrsim 99\%$ and $\gtrsim 95\%$ confidence level
respectively in both \nhi~and \bhi~distributions. The KS test gives no
significant difference between the $-2\sigma$ and $-1\sigma$ regions in
both \nhi~and \bhi~distributions (see Table \ref{tab:4}). These results
do not change significantly when correcting \gimic~to match the
observed \nhi~distribution using a different $\langle F \rangle$ and
$S/N$ values. We do not attempt to make a more detailed comparison
between distributions coming from real data (void, void-edge samples)
and the different \gimic~regions as there are already known differences
between them (see \S \ref{gimic:properties}).

\section{Discussion}\label{discussion}

\subsection{Three \lya~forest populations}\label{discussion:lya}
Our first result is that there is a $>99\%$ c.l. excess of \lya~systems
at the edges of galaxy voids compared to a random distribution (see \S
\ref{gasaroundvoids} and Figure \ref{fig:X_D_dist}). Our random sample
was normalized to have the same density of systems in the whole
volume. Then, an excess in a sub-volume means necessarily a deficit in
another. Given that we found no significant difference in the number of
systems in voids with respect to the random expectation, the excess is
not explained by a deficit of \lya~systems inside galaxy voids. The
observed excess seems more related to the lack of systems found at
distances $\gtrsim5$ \mpc~outside the cataloged voids (and
viceversa). Thus, despite the fact that we see \lya~clustered at the
edges of galaxy voids, it is not clear from this data that \lya-voids
at low-$z$ exist at all (see \citealt{carswell87} for similar result at
high-$z$; although see \citealt{williger00}). This picture is somewhat
different from the case of galaxies, where galaxy voids are present
even in the distribution of low mass galaxies
\citep[e.g.,][]{peebles01,tikhonov09}. There is agreement though in the
sense that both \lya~system and galaxy distributions have their peaks
at the edges of galaxy voids. We observe a typical scale length of the
excess to be $\sim 5$ \mpc, consistent with numerical predictions for
the typical radius of the filamentary structure of the `cosmic web'
\citep[$\sim 2$ \mpc; see][]{gonzalez10,aragon-calvo10, bond10}. Note
that this scale length is approximately twice the scale associated with
a velocity uncertainty of $\Delta v \approx 200$ \kms~at $z=0$. Such
dispersions could be present in our void-edge sample.

In our data analysis we have defined three samples of absorption
systems, based on how they are located with respect to the closest
galaxy void (see \S \ref{definition}). Let us consider now a very
simple model in which we have only two LSS environments: under-dense
and over-dense LSS. Then we could relate all the `random-like'
\lya~forest systems found in the void sample ($X<0.9$ and/or $D<-2$
\mpc) with the under-dense LSS, while all the systems associated with
the excess over random ($0.9\le X < 1.3$ and/or $-2\le D < 4$ \mpc) to
the over-dense LSS\footnote{We have left the unclassified systems out
  of this interpretation.}. Note that the fact that we cannot
distinguish between the under-dense LSS \lya~distribution and a random
distribution does not mean that the former is really random. If these
\lya~forest systems follow the underlying dark matter distribution
\citep[e.g., see][]{croft98}, they should have a non-negligible
clustering amplitude that is not observed only because of the lack of
statistical power of our sample. In relative terms, considering the
under-dense LSS \lya~as random seems to be a good approximation though,
especially for the large scale distances involved in this work ($> 1$
Mpc). This is also supported by the very low auto-correlation amplitude
observed in the whole population of \lya~forest systems at such scales
\citep[e.g.,][]{croft98,rollinde03,crighton11}. This `random' behavior
of \lya~systems in the under-dense LSS can be understood as originating
in structures still evolving from the primordial density perturbations
in the linear regime. At $z=0$ however, the majority of the mass
resides at the edges of voids (in the `cosmic web') whose density
perturbations have reached non-linear evolution regime at higher
redshifts.  For reference, we expect the under-dense and over-dense LSS
to have typical $\delta \gtrsim 0$ and $\delta \lesssim 0$
respectively, where $\delta$ is the density contrast defined as,

\begin{equation}
$$\delta \equiv \frac{\rho-\langle\rho\rangle}{\langle \rho \rangle}
  \ \ \rm{,}$ $
\label{eq:density}
\end{equation}

\noindent where $\rho$ is the density and $\langle\rho\rangle$ is the
mean density of the universe. Note however that these LSS environments
are not defined by a particular density but rather by a topology
(voids, walls, filaments).

Theoretical arguments point out that the observed column density of
neutral hydrogen at a fixed $z$ is,

\begin{equation}
N_{\rm HI} \propto \rho_{\rm H}^{1.5} \ T^{-0.26} \ \Gamma^{-1}
\ f_g^{0.5} \ \ \rm{,}
\label{eq:nhi}
\end{equation}

\noindent where $\rho_{\rm H}$ is the density of hydrogen, $T$ is the
temperature of the gas, $\Gamma$ is the hydrogen photo-ionization rate
and $f_g$ is the fraction of mass in gas \citep{schaye01}. In the
diffuse IGM it has been predicted that $T \propto \rho_{\rm
  H}^{\alpha}$, where $\alpha \approx 0.59$ \citep{hui97}. This implies
that for a fixed $\Gamma$, the main dependence of \nhi~is due to
$\rho_{\rm H}$ as \nhi~$\propto \rho_{\rm H}^{1.4}$. Then, despite the
extremely low densities inside galaxy voids we can still observe
\lya~systems, although only the ones corresponding to the densest
structures.

Let us consider the predicted ratio between \nhi~observed inside voids
and at the edges of voids as,

\begin{equation}
\frac{N_{\rm HI}^{\rm void}}{N_{\rm HI}^{\rm edge}} \approx \left(
\frac{\rho_{\rm H}^{\rm void}}{\rho_{\rm H}^{\rm edge}} \right)^{1.4}
\left( \frac{\Gamma^{\rm void}}{\Gamma^{\rm edge}}\right)^{-1} \left(
\frac{f_g^{\rm void}}{f_g^{\rm edge}} \right)^{0.5} \ \ \rm{.}
\label{eq:nhi_frac}
\end{equation}

\noindent Given that the timescale for photons to travel along $\sim
10-20$ \mpc~is $\ll 1$ Gyr, we can consider $\Gamma^{\rm void} \approx
\Gamma^{\rm edge}$. Even if we assume that the gas inside voids has not
formed galaxies, $f_g^{\rm void} \gtrsim f_g^{\rm edge}$, because $f_g$
is dominated by the dark matter. This implies that a given observed
\nhi~inside and at the edge of galaxy voids will correspond to similar
densities of hydrogen ($\rho^{\rm void}_{\rm H} \approx \rho^{\rm
  edge}_{\rm H}$). This is important because it means that the
\lya~forest in the under-dense LSS is not different than the over-dense
LSS one, and two systems with equal \nhi~are comparable, independently
of its large scale environment.

If there were no galaxies, this simple model may suffice to explain the
differences in the observed \lya~population. The fact that some of the
\lya~systems are directly associated with galaxies cannot be neglected
though. There is strong evidence from observations
\citep[e.g.,][]{lanzetta95,chen98,stocke06,morris06,chen09,crighton11,prochaska11,rakic11,rudie12}
and simulations \citep[e.g.,][]{fumagalli11,stinson11} that $N_{\rm
  HI}\gtrsim 10^{15}$ \cm~systems are preferentially found within a
couple of hundred kpc of galaxies. Probably an appropriate
interpretation of such a result is that galaxies are always found in
`local' ($\lesssim 100$ $h^{-1}$kpc) high \nhi~density regions. Then, a
plausible scenario would require {\it at least} three types of
\lya~forest systems: (1) containing embedded galaxies, (2) associated
with over-dense LSS but with no close galaxy and (3) associated with
under-dense LSS but with no close galaxy. For convenience, we will
refer to the first type as `halo-like', although with the caution that
these systems may not be gravitationally bound with the galaxy.

Given that there are galaxies inside galaxy voids, the `halo-like'
\lya~systems will be present in both low and high density LSS
environments (galaxies are a `local' phenomenon). The contribution of
the `halo-like' in galaxy voids could be considered small
though. Assuming this contribution to be negligible, we can estimate
the fraction of \lya~systems in the under-dense LSS as $\approx 25-30\%
\pm 6\%$\footnote{These numbers come from $28\pm5 (22\pm5)$ and $61\pm8
  (65\pm8)$ systems found at $X<0.9$ ($D<-2$ \mpc) and at $0.9\le
  X<1.3$ ($-2 \le D <4$ \mpc) respectively (see \S
  \ref{gasaroundvoids}).}. Likewise, $\approx 70-75\% \pm 12\%$\fnn~of
the \lya~forest population are due to a combination of systems
associated with galaxies and systems associated with the over-dense
LSS. We could estimate the contribution of `halo-like' absorbers by
directly looking for and counting galaxies relatively close to the
absorption systems. A rough estimation can be done by assuming that
galaxy halos will have only $N_{\rm HI} \ge 10^{14}$ \cm~systems,
leading to a contribution of $\approx 12-15\% \pm 4\%$\footnote{From
  either $13/89$ (excluding the unclassified sample) or $13/106$
  (including the unclassified sample). We have assumed Poisson
  uncertainty.} in our sample.

In summary, our results require {\it at least} three types of
\lya~systems to explain the observed \lya~forest population at low-$z$
(\nhi~$\gtrsim10^{12.5}$ \cm):

\begin{itemize}
\item {\it Halo-like:} \lya~with embedded nearby galaxies ($\lesssim
  100$ $h^{-1}$kpc) and so directly correlated with galaxies ($\approx
  12-15\% \pm 4\%$),
\item {\it Over-dense LSS:} \lya~associated with the over-dense LSS
  that are correlated with galaxies only because both populations lie
  in the same LSS regions ($\approx 50-55\% \pm 13\%$) and,
\item {\it Under-dense LSS:} \lya~associated with the under-dense LSS
  with very low auto-correlation amplitude that are not correlated with
  galaxies ($\approx 25-30\% \pm 6\%$).
\end{itemize}

\noindent The relative contributions of these different
\lya~populations is a function of the lower \nhi~limit. Low
\nhi~systems dominate the \lya~column density distribution. Then, given
that under-dense LSS \lya~systems tend to be of lower column density
than the other two types, we expect the contribution of `random-like'
\lya~to increase (decrease) while observing at lower (higher)
\nhi~limits. Note that there are not sharp \nhi~limits to differentiate
between our three populations (see Figure \ref{fig:2}). The `halo-like'
is defined by being close to galaxies while the `LSS-like' ones are
defined in terms of a LSS topology (voids, wall, filaments).

Motivated by a recently published study on the \lya/galaxy association
by \citet{prochaska11}, we can set a conservative upper limit to the
`halo-like' contribution. These authors have found that nearly all
their observed $L \ge 0.01L^*$ galaxies ($33/37$) have \nhi~$\ge
10^{13.5}$ \cm~absorption at impact parameters $<300$
$h^{-1}_{72}$kpc. If we invert the reasoning and assume an extreme
(likely unrealistic) scenario where {\it all} the \nhi~$\ge 10^{13.5}$
\cm~are directly associated with galaxies, then, the `halo-like'
contribution will have an upper limit of $< 33\% \pm 7\%$\footnote{From
  either $29/89$ (excluding the unclassified sample) or $35/106$
  systems in our sample (including the unclassified sample). We have
  assumed Poisson uncertainty.}. Consequently, the contribution of the
over-dense LSS to the \lya~population will be $> 37-42\% \pm
14\%$. Still, note that we have found several systems with $10^{13.5}
\lesssim N_{\rm HI} \lesssim 10^{14.5}$ \cm~inside galaxy voids for
which a direct association with galaxies is dubious (see Figure
\ref{fig:2}). Also note that only $\sim 10\%$ ($ \sim 0\%$) of the
\lya~systems between $10^{13.5}<N_{\rm HI}< 10^{14.5}$ \cm~may be
associated with a galaxy at impact parameters $<300$ kpc ($<100$ kpc)
in the \citet{prochaska11} sample (see their figure 4).

Our findings are consistent with previous studies pointing out a
non-negligible contribution of `random' \lya~systems (at a similar
\nhi~limit) of $\approx 20-30\%$ \citep{mo94,stocke95,penton02}. These
authors estimated that $\approx 70-80\%$ of the \lya~population is
associated with either LSS (galaxy filaments) or galaxies. Note that
\citet{mo94} put a upper limit of $\approx 20\%$ being directly
associated with galaxies, which is also consistent with our
estimation. Our result is also in accordance with the previous
estimation that $22\% \pm 8\%$ (\citealt{penton02}; based on $8$
systems) and $17\% \pm 4\%$ (\citealt{wakker09}; based on $17$ systems)
of the \lya~systems lie in voids (defined as locations at $>3$
$h^{-1}_{70}$ Mpc from the closest $>L^*$ galaxy). This is in contrast
with early models that associated {\it all} \lya~systems with galaxies
\citep[e.g.,][]{lanzetta95,chen98}.

Although there is general agreement with recently proposed models to
explain the origin of the low-$z$ \lya~forest
\citep[e.g.,][]{wakker09,prochaska11}, we emphasize that our
interpretation is qualitatively different and adds an important
component to the picture: the presence of the under-dense LSS
(`random-like') systems. For instance, assuming infinite filaments of
typical widths of $\approx 400$ $h^{-1}_{72}$kpc around galaxies,
\citet{prochaska11} argued that {\it all} \lya~systems at low-$z$
belong either to the circum-galactic medium (CGM\footnote{According to
  the \citet{prochaska11} definition, {\it the CGM corresponds to
    highly ionized medium around galaxies at distances greater than the
    virial radius but smaller than $\sim$ 300 kpc, that need not be
    causally connected (associated, gravitationally bound) with these
    galaxies}. We do not see a clear advantage of adopting this
  terminology and so we use `IGM' instead to refer to the same
  medium. We only make a distinction between LSS (voids, walls,
  filaments) and the ones with embedded nearby galaxies ($\lesssim 100$
  $h^{-1}$kpc).}; which includes our `galaxy halo' definition) or the
filamentary structure in which galaxies reside (equivalent to our
over-dense LSS definition). Our findings are not fully consistent with
this hypothesis, as neither the `CGM model' nor the `galaxy filament
model' seem likely to explain the majority of our under-dense LSS
absorbers at \nhi~$\lesssim10^{13.5}$ \cm. To do so there would need to
be a whole population of unobserved galaxies (dwarf spheroidals?)
inside galaxy voids with an auto-correlation amplitude as low as the
`random-like' \lya~one. As discussed by \citealt{tikhonov09}, very low
surface brightness dwarf spheroidals could be a more likely explanation
than dwarf irregulars because the latter should have been observed with
higher incidences in recent \hi~emission blind surveys inside galaxy
voids (e.g., HIPASS, \citealt{doyle05}). On the other hand, the
formation of dwarf spheroidals inside galaxy voids is difficult to be
explained from the current galaxy formation paradigm (see
\citealt{tikhonov09} for further discussion). As mentioned, it seems
more natural to relate the majority of the under-dense LSS absorbers
with the peaks of extremely low density structures inside galaxy voids,
still evolving linearly from the primordial density perturbations that
have not formed yet galaxies because of their low densities. Our
interpretation can be tested by searching for galaxies close to our
lowest \nhi~void-absorbers (see Figure \ref{fig:2}). Another prediction
of our interpretation is that the vast majority of
\nhi~$\lesssim10^{13}$ \cm~systems should reside inside galaxy
voids. If the QSO sightlines used here were observed at higher
sensitivities, weak \lya~systems should preferentially appear at $X<1$
($D<0$ \mpc). Therefore, we should expect to have an {\it
  anti}-correlation between \nhi~$\lesssim10^{13}$ \cm~and galaxies.

\subsection{\nhi~and \bhi~distributions}

Our second result is that there is a systematic difference ($\gtrsim
98\%$ c.l.) between the column density distributions of \lya~systems
found within, and those found at the edge of, galaxy voids. Void
absorbers have more low column density systems than the void-edge
sample (see Figure \ref{fig:3}). A similar trend is found in \gimic,
where low density environments present smaller \nhi~values than higher
density ones (see Figure \ref{fig:8}, left panel). This can be
explained by the fact that baryonic matter follows the underlying dark
matter distribution. Then, the highest density environments should be
located at the edges of voids (in the intersection of walls and
filaments), consequently producing higher column density absorption
than in galaxy voids \citep[e.g., see][]{schaye01}.

Also, by construction, there is a higher chance to find galaxies at the
edges rather than inside galaxy voids. Assuming that some of the
\lya~forest are associated with galaxy halos (see \S
\ref{discussion:lya} for further discussion), then this population
should be present mainly in our void-edge sample. As galaxy halos
correspond to local density peaks, we should also expect on average
higher column density systems in this population. Given that galaxies
may affect the properties of the surrounding gas, there could be
processes that only affect \lya~systems close to galaxies. For
instance, the distribution of \lya~systems around galaxy voids seems to
show a two-peaked shape (see Figure \ref{fig:X_D_dist}). We speculate
that this could be a signature of neutral hydrogen being ionized by the
ultra-violet background produced by galaxies \citep[see
  also][]{adelberger03}, mostly affecting \nhi~$\lesssim 10^{13}$
\cm~inside the filamentary structure of the `cosmic web'. Another
explanation could be that in the inner parts of the filamentary
structure, \lya~systems get shock heated by the large gravitational
potentials, raising their temperature and ionization state
\citep[e.g.,][]{cen99}. A third possibility is that it could be a
signature of bulk outflows as the shift between peaks is consistent
with a $\Delta v \approx 200-300$ \kms. On the other hand, the two
peaks could have distinct origins as the first one may be related to an
excess of \nhi~$\lesssim 10^{13}$ \cm~systems, probably associated with
the over-dense LSS in which galaxies reside, while the second one may
be related to an excess of \nhi~$\gtrsim 10^{14}$ \cm~systems, more
likely associated with systems having embedded galaxies. As mentioned,
we cannot prove the reality of this two-peaked signature at a high
confidence level from the current sample and so we leave to future
studies the confirmation or disproof of these hypotheses.

The \gimic~data analysis shows a clear differentiation of
\bhi~distributions in different density environments (see Figure
\ref{fig:8}, right panel). Low density environments have smaller
\bhi~values than higher density ones. We see a similar trend in the
real data between our void and void-edge absorber samples, although
only at a $\gtrsim90\%$ of confidence level (i.e., not very
significant; see Figure \ref{fig:4}). The main mechanisms that
contribute to the observed line broadening are temperature, local
turbulence and bulk motions of the gas (excluding systematic effects
from the line fitting process or degeneracy with \nhi~for saturated
lines). Naturally, in high density environments, we would expect to
have greater contributions from both local turbulence and bulk motions
compared to low density ones. The gas temperature is also expected to
increase from low density environments to high density ones. As
previously mentioned, theoretical arguments predict the majority of the
diffuse IGM will have temperatures related to the density by $T \propto
\rho^{\alpha}$ with $\alpha>0$ \citep{hui97,theuns98,schaye99}. This is
also seen in density-temperature diagrams drawn from current
hydrodynamical cosmological simulations
\citep[e.g.,][]{dave10,tepper-garcia12}. Therefore, our findings are
consistent with current expectations.

\subsection{Future work}

The high sensitivity of the recently installed {\it Cosmic Origins
  Spectrograph} \citep[COS/HST,][]{green12} in the UV (especially the
far-UV), will allow us to improve the \nhi~completeness limit compared
with current surveys. This will considerably increase the number of
observed \lya~absorption systems at low-$z$. In the short term, there
are several new QSO sightlines scheduled for observations (or already
observed) with COS/HST that intersect the SDSS volume. Combining these
with current and future galaxy void catalogs, we expect to increase the
statistical significance of the results presented in this work. COS/HST
will also allow observations of considerably more metal lines
(especially \ovi) than current IGM surveys. Again, in combination with
LSS surveys, this will be very useful for studies on metal enrichment
in different environments. For instance, we have identified $8$ systems
with observed \ovi~absorption from STIS/HST in our sample. Three of
these lie inside voids at $X\approx \{0.6,0.7,0.9\}$ ($D\approx
\{-5.4,-4.6,-1.6\}$ \mpc) respectively. The first two systems that lie
inside voids correspond to the highest \nhi~values (\nhi $> 10^{14.5}$
\cm; see Figure \ref{fig:2}). We have performed a search in the SDSS
DR8 for galaxies in a cylinder of radius $1$ $h^{-1}_{71}$ Mpc and
within $\pm 200$ \kms~around these $2$ absorbers (both systems belong
to the same sightline and are at a similar redshift; one of them shows
\civ~absorption also). We found $9$ galaxies with these constraints,
hinting on a possible association of these systems with a void
galaxy. The one at the very edge of the void limit has \nhi~$=10^{13.14
  \pm 0.07}$ \cm~and $N_{\rm OVI}=10^{13.69\pm0.18}$ \cm, and it could
in principle be associated with the over-dense LSS. The other $5$
\ovi~absorbers lie in our void-edge sample and have \nhi~$>10^{13.5}$
\cm, so they are likely to be associated with galaxies.  None of the
observed \ovi~lie in our unclassified sample. The current sample of
\ovi~systems is very small, and so we do not aim to draw statistical
conclusions from them. However, these systems individually offer
interesting cases worth further investigation. We intend to perform a
carefully search for galaxies that could be associated to each of the
\lya~absorbers presented in our sample in future work. In the longer
term, it will be possible to extend similar analysis to well defined
galaxy filaments and clusters when the new generation of galaxy surveys
are released.

A scenario with three different types of \lya~forest systems, as
proposed here, can help to interpret recent measurements of the
cross-correlation between \lya~and galaxies
\citep{chen05,ryan-weber06,wilman07,chen09,shone10,rudie12}. These
studies come mainly from pencil beam galaxy surveys around QSO
sightlines where identifying LSS such as voids or filaments is more
challenging.  As mentioned, different \lya~systems are not separated by
well defined \nhi~limits and so we suggest using our results to
properly account for under-dense LSS (`random-like') absorbers in
gas/galaxy cross-correlations. Truly random distributions are easy to
correct for, as they lower the amplitude of the correlations at all
scales. Then, acknowledging these `random-like' absorbers, it will be
possible to split the correlation power in its other two main
components: gas in galaxy halos and gas in the over-dense LSS. Our
group is currently working in a future paper to study the gas/galaxy
cross-correlation, in which these corrections will be taken into
account.

\section{Summary}\label{summary}
We have presented a statistical study of \hi~\lya~absorption systems
found within and around galaxy voids at $z\lesssim0.1$. We found a
significant excess ($>99\%$ c.l.) of \lya~systems at the edges of
galaxy voids with respect to a random distribution, over a $\sim 5$
\mpc~scale. We have interpreted this excess as being due to
\lya~systems associated with both galaxies (`halo-like') and the
over-dense LSS in where galaxies reside (the observed `cosmic web'),
accounting for $\approx 70-75\% \pm 12\%$ of the \lya~population. We
found no significant difference in the number of systems inside galaxy
voids compared to the random expectation. We therefore infer the
presence of a third type of \lya~systems associated to the under-dense
LSS with a low auto-correlation amplitude ($\approx$ random) that are
not associated with luminous galaxies. These `random-like' absorbers
are mainly found in galaxy voids. We argue that these systems can be
associated with structures still growing linearly from the primordial
density fluctuations at $z=0$ that have not yet formed galaxies because
of their low densities. Although the presence of a `random' population
of \lya~absorbers was also inferred (or assumed) in previous studies,
our work presents for the first time a simple model to explain it (see
\S \ref{discussion:lya}). Above a limit of \nhi~$\gtrsim 10^{12.5}$
\cm, we estimate that $\approx 25-30\% \pm 6\%$ of \lya~forest systems
are `random-like' and not correlated with luminous galaxies. Assuming
that {\it only} \nhi~$\ge 10^{14}$ \cm~systems have embedded galaxies
nearby, we have estimated the contribution of the `halo-like'
\lya~population to be $\approx 12-15\% \pm 4\%$ and consequently
$\approx 50-55\% \pm 13\%$ of the \lya~systems to be associated with
the over-dense LSS.

We have reported differences between both the column density (\nhi) and
the Doppler parameter (\bhi) distributions of \lya~systems found inside
and at the edge of galaxy voids observed at the $>98\%$ and $>90\%$ of
confidence level respectively. Low density environments (voids) have
smaller values for both \nhi~and \bhi~than higher density ones (edges
of voids). These trends are theoretically expected. We have performed a
similar analysis using simulated data from \gimic, a state-of-the-art
hydrodynamical cosmological simulation. Although \gimic~did not give a
perfect match to the observed column density distribution, the
aforementioned trends were also seen. Any discrepancy between
\gimic~and real data could be due to low number statistic fluctuations
and/or a combination of several observational effects.

In summary, our results are consistent with the expectation that the
mechanisms shaping the properties of the \lya~forest are different in
different LSS environments. By focusing on a `large scale' ($\gtrsim10$
Mpc) point of view, our results offer a good complement to previous
studies on the IGM/galaxy connection based on `local' scales ($\lesssim
2$ Mpc).

\section*{Acknowledgments}
We thank the anonymous referee for helpful comments which improved the
paper. We thank Charles Danforth for having kindly provided $S/N$
information for the QSO spectra used in this work. N.T. acknowledges
grant support by CONICYT, Chile (PFCHA/{\it Doctorado al Extranjero
  1$^{\rm a}$ Convocatoria}, 72090883).

\bibliographystyle{mn2e_trunc8}
\bibliography{LSS-abs}

\appendix

\section[]{Check for systematic effects}\label{systematic}
In this section we investigate possible biases or systematic effects
that could be present in the data analysis.

\subsection{Comparison between our subsample and the whole DS08 sample}\label{reference}
In this section we explore whether the subsample of \hi~systems used
here is statistically different from the rest of the \hi~population
found in the other sightlines of the DS08 catalog. For this, we
compared the column density (\nhi) and Doppler parameter (\bhi)
distributions between the $106$ absorption systems found inside the
void catalog volume ($0.01 \le z_{\rm abs} \le 0.102$ in $11$
sightlines) with those $545$ outside this volume. The KS test gives a
probability of $\approx 5\%$ and $\approx 90\%$ that both \nhi~and
\bhi~distributions, inside and outside the void catalog volume, are
drawn from the same parent distribution respectively. This shows there
is no significant difference in the \bhi~distribution and an $\approx
2\sigma$ difference between the \nhi~distributions.

One explanation for such a difference could be due to an intrinsic
evolution of the \lya~forest between $z_{\rm abs} \lesssim 0.102$ and
$0.102 \lesssim z_{\rm abs} <0.4$. To check this, we performed a KS
comparison between the $286$ absorbers at $z_{\rm abs}<0.102$
(regardless of whether they are inside the void catalog volume or not)
with the $365$ systems at higher redshifts, to look for a possible
difference in both \nhi~and \bhi~distributions. The distribution of
\bhi~does not show any difference (KS Prob. $\approx 87\%$). On the
other hand, the KS probability for the two \nhi~distributions is
$\approx 8\%$, hinting that such evolution could be present. We note,
however, that an observational bias between low and high-$z$ systems
(e.g., due to different selection functions) could also explain the
observed difference. We did not further explore this matter.

Finally, we repeated the previous comparisons between the $106$ systems
inside the void catalog volume with those $155$ outside it having $0.01
\le z_{\rm abs} \le 0.102$. The KS test gives this time a probability
of $\approx 23\%$ and $\approx 81\%$ of both \nhi~and \bhi~inside and
outside the void catalog volume are drawn from the same parent
distribution, respectively. No significant ($>2\sigma$) differences are
found for these samples. Thus, we conclude that the properties of the
systems in the sightlines used in this work are not statistically
different from the properties of the systems in other sightlines when
we restrict ourselves to the same redshift range.

We note that for our comparisons we use systems with \nhi~below the
completeness of the column density distribution itself ($N_{\rm HI}
\lesssim 10^{13.4}$ cm$^{-2}$). The classification of absorber
environment does not depend on column density but rather corresponds to
a geometrical association of absorbers with known galaxy
voids. Therefore, this should not affect any of our results. We also
checked that none of the AGN used here were observed in particular for
having sightlines intersecting a known void region.

\begin{figure}
  \includegraphics[width=90mm]{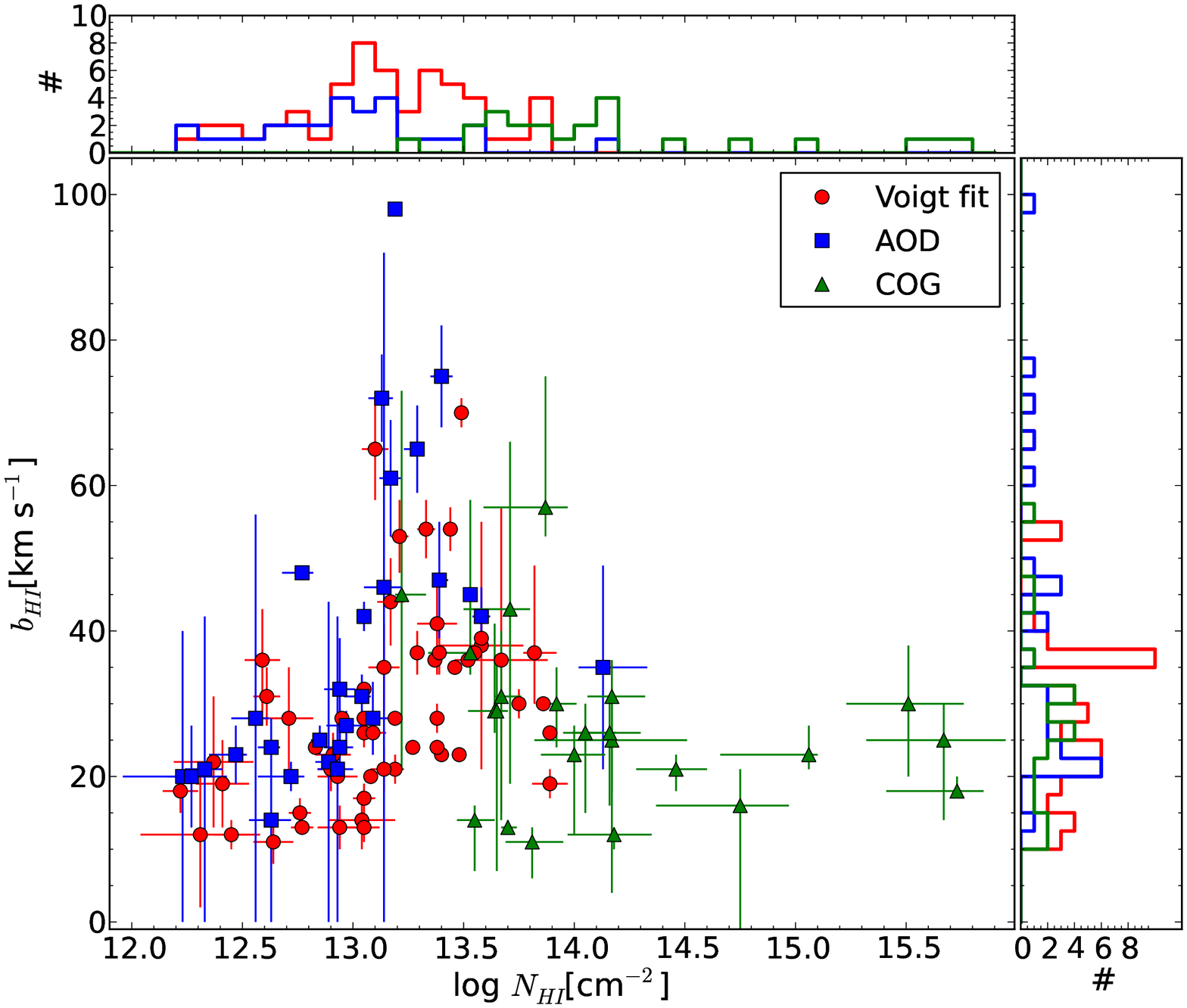}
  \caption{\hi~column density versus \hi~Doppler parameter for systems
    in our sample. Different characterization methods (see \S
    \ref{systematic:method}) are shown by different colors/symbols.
    Red circles correspond to systems measured by Voigt profile
    fits. Blue squares correspond to systems measured by the AOD
    method. Green triangles correspond to systems measured by a COG
    solution. Histograms (arbitrary binning) are also shown around the
    main panel using the previous color definition.}\label{fig:5}

\end{figure}

\subsection{Systematics in the DS08 characterization method}\label{systematic:method}
A particular source of concern is the fact that DS08 used different
methods for obtaining the \hi~column densities and Doppler parameters
in the catalog. They used curve-of-growth (COG) solutions when other
Lyman series lines were available other than \lya. For the rest of the
systems they used either a single Voigt profile fit (preferentially for
strong or blended lines) and/or the apparent optical depth method
\citep[AODM,][preferentially for weak, asymmetric or noisy
  lines]{savage91}. We will refer to these as \lya-only methods. Given
that we have been using systems without distinction between these
different methods we will explore any possible bias that this could
produce.

Figure \ref{fig:5} shows \hi~column densities versus \hi~Doppler
parameters for the $106$ systems in our sample, where the different
characterization methods used by DS08 are shown in different
colors/symbols. It is possible to observe a clear separation between
column densities derived from \lya-only methods and COG solutions. This
difference is a direct consequence of the fact that COG solutions can
only be obtained for systems showing a high order Lyman series line,
whose equivalent widths are smaller than \lya~for the same \nhi~and
\bhi~values (because they have smaller wavelengths and smaller
oscillator strengths). This results in a shift of the completeness
level to higher column densities for the COG solutions, as can be seen
in Figure \ref{fig:5}. On the other hand, no significant difference
seems to be present for \bhi~values between different characterization
methods, which has been confirmed by a KS test.

Our definitions of the different LSS samples do not depend on either
\nhi~nor \bhi~but it rather correspond to a geometrical selection of
how far away a system lies from the center of a void. Thus, the
aforementioned difference observed in \nhi~should not affect the
conclusions drawn from this study as long as our different LSS samples
have similar contributions of systems measured with one method or the
other. Assuming a Poisson distribution for the number of systems in
each sample, there is a relative contribution of $82\% \pm 18\%$ ($82\%
\pm 18\%$)\fn, $73\% \pm 17\%$ ($74\% \pm 17\%$)\fn~and $82\% \pm 30\%$
($79\% \pm 27\%$)\fn~for systems measured by either Voigt fit or AOD
methods in the void, void-edge and unclassified samples
respectively. These numbers are all consistent with each other within
$1\sigma$. Therefore we conclude that different characterization
methods do not introduce an important bias in our analysis.

There is also a reported systematic effect seen when using \lya~only
systems by which \nhi~estimates are typically (with large scatter) low
by a factor of $\sim 3$ while \bhi~are high by a factor of $\sim 2$
\citep{shull00,danforth06}. This systematic effect is important for
higher column densities. Our sample is dominated by systems with
\nhi$<10^{14}$ \cm, for which the effect is smaller than the quoted
numbers \citep[see figures 2 and 3 from][]{danforth06}. So, despite the
fact that our sample is dominated by \lya-only measurements, we do not
consider this effect to be important. Finally, as previously argued,
even if any of these effects are present they will affect each of our
LSS samples in roughly the same proportion.  \begin{figure}
  \includegraphics[width=90mm]{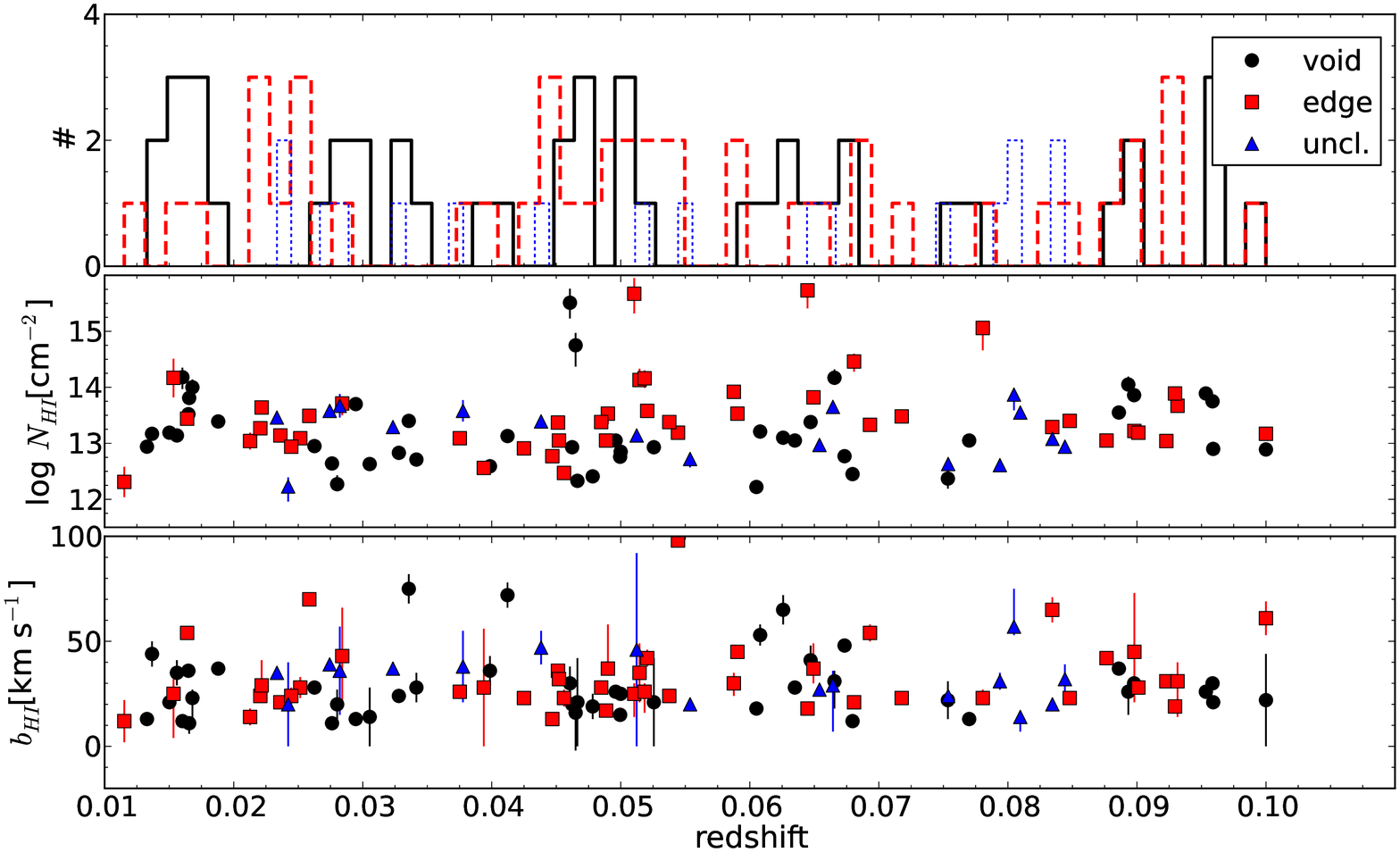}
  \caption{Distribution of observables as a function of
    redshift. Middle and bottom panels show the distribution of
    \nhi~and \bhi~respectively. Different LSS samples are shown by
    different color/symbols: void (black circles), void-edge (red
    squares) and unclassified (blue triangles). Top panel shows the
    distribution of different LSS as a function of redshift: void
    (black-solid line), void-edge (red-dashed line) and unclassified
    (blue-dotted line). For simplicity we only show LSS definitions
    based on $X$.}  \label{fig:6}

\end{figure}

\subsection{Observables as a function of redshift}
We also checked that no bias is present in our samples as a function of
redshift. Figure \ref{fig:6} shows both \nhi~and \bhi~values as a
function of redshift (middle and bottom panels). Systems belonging to
different LSS samples are shown by different color/symbols. Both
distributions look very uniform across the full redshift range. The top
panel shows the distribution of void, void-edge and unclassified
absorbers as a function of redshift. The KS test shows no significant
difference between these LSS samples. We conclude that there is no
evident systematic effect as a function of redshift. For simplicity we
have only used LSS definitions based on $X$ in Figure \ref{fig:6}, but
the previous results also hold using $D$ instead.

The catalog of P12 used a nearly complete, magnitude limited sample of
galaxies to define the voids.  Despite this, we performed an
independent check by looking at the mean radius $\langle R_{\rm void}
\rangle$ as a function of redshift. If the catalog is well defined, we
should expect to have this radius constant across redshift range
(assuming no measurable evolution). We confirmed that this is actually
the case by dividing the sample in $6$ redshift bins and measuring the
mean value. We found that the mean radius is constant with $\langle
R_{\rm void} \rangle \approx 13 \pm 3$ \mpc~in each bin.


\bsp
\label{lastpage}
\end{document}